\documentclass[compsoc,conference,a4paper,10pt]{IEEEtran}
\pdfoutput=1
\usepackage{hyperref}       % hyperlinks
\usepackage{url}            % simple URL typesetting
\usepackage{booktabs}       % professional-quality tables
\usepackage{nicefrac}       % compact symbols for 1/2, etc.
\usepackage{microtype}      % microtypography
\usepackage{graphicx}
\usepackage{doi}
\usepackage{color}
\usepackage{xcolor}
\usepackage{cite}
\usepackage{colortbl}
\usepackage[font=footnotesize]{caption}
\usepackage{multicol}
\usepackage{amsmath,amsfonts,stmaryrd,xspace,enumitem}
\usepackage{tablefootnote}
\usepackage{multirow}
\usepackage{algorithmic,algorithm}
\usepackage{subfigure}
\usepackage{verbatim}
\usepackage{wrapfig}
\usepackage{tikz}
\usepackage{circledsteps}
\usepackage{makecell}
\usepackage{pifont}
\usepackage{svg}
\usepackage{setspace}
\usetikzlibrary{shapes.geometric, positioning}

\def\etc{\textit{etc.}}
\def\eg{\textit{e.g.}}
\def\etal{\textit{et~al.}}
\def\aka{\textit{a.k.a.}}

\def\chatiot{\textsc{ChatIoT}} 
\def\iotrgen{\textbf{{IoT-RG}}}
\def\datakit{\textbf{{DataKit}}} 

\newcommand{\centered}[1]{\begin{tabular}{c} #1 \end{tabular}}
\definecolor{verylightgray}{gray}{0.95}

\definecolor{keywordcolor}{rgb}{0.5,0.1,0.6} % Purple keywords
\definecolor{stringcolor}{rgb}{0.2,0.6,0.2}  % Green strings
\definecolor{commentcolor}{rgb}{0.5,0.5,0.5} % Gray comments
\definecolor{bgcolor}{rgb}{0.95,0.95,0.95}   % Light gray background for the code

\def\BibTeX{{\rm B\kern-.05em{\sc i\kern-.025em b}\kern-.08em
    T\kern-.1667em\lower.7ex\hbox{E}\kern-.125emX}}
\begin{document}

\title{\chatiot: Large Language Model-based Security Assistant for Internet of Things with Retrieval-Augmented Generation
}

%\author{\IEEEauthorblockN{Anonymous Authors}}

\author{
    \IEEEauthorblockN{Ye Dong}
    \IEEEauthorblockA{Singapore University of Technology and Design\\
    ye\_dong@sutd.edu.sg}\\
    \IEEEauthorblockN{Sudipta Chattopadhyay}
    \IEEEauthorblockA{Singapore University of Technology and Design\\
    sudipta\_chattopadhyay@sutd.edu.sg}
    \and 
   \IEEEauthorblockN{Yan Lin Aung}
    \IEEEauthorblockA{Singapore University of Technology and Design\\
    linaung\_yan@sutd.edu.sg}\\
    \IEEEauthorblockN{Jianying Zhou}
    \IEEEauthorblockA{Singapore University of Technology and Design\\
    jianying\_zhou@sutd.edu.sg}
}

\maketitle

\begin{abstract}
Internet of Things (IoT) has gained widespread popularity, revolutionizing industries and daily life. However, it has also emerged as a prime target for attacks.
Numerous efforts have been made to improve IoT security, and substantial IoT security and threat information, such as datasets and reports, have been developed.
However, existing research often falls short in leveraging these insights to assist or guide users in harnessing IoT security practices in a clear and actionable way.
In this paper, we propose \chatiot, a large language model (LLM)-based IoT security assistant designed to disseminate IoT security and threat intelligence.
By leveraging the versatile property of retrieval-augmented generation (RAG), \chatiot\ successfully integrates the advanced language understanding and reasoning capabilities of LLM with fast-evolving IoT security information.
Moreover, we develop an end-to-end data processing toolkit to handle heterogeneous datasets.
This toolkit converts datasets of various formats into retrievable documents and optimizes chunking strategies for efficient retrieval.
Additionally, we define a set of common use case specifications to guide the LLM in generating answers aligned with users' specific needs and expertise levels.
Finally, we implement a prototype of \chatiot\ and conduct extensive experiments with different LLMs, such as LLaMA3, LLaMA3.1, and GPT-4o. Experimental evaluations demonstrate that \chatiot\ can generate more reliable, relevant, and technical in-depth answers for most use cases.
When evaluating the answers with LLaMA3:70B, \chatiot\ improves the above metrics by over $10\%$ on average, particularly in relevance and technicality, compared to using LLMs alone.
\end{abstract}


\section{Introduction}\label{sec:intro}
Internet of Things (IoT) constitutes a vast network of physical and virtual entities, characterized by their sensing and/or actuation capabilities, programmability features, and unique identifiers. 
This interconnected infrastructure has rapidly expanded, and IoT connections surpassed non-IoT connections in 2020. 
It is estimated that over 40 billion IoT devices will be integrated into homes and workplaces via sensors, processors, and software by 2030\footnote{\scriptsize \url{https://iot-analytics.com/number-connected-iot-devices/}}.
With the rapid developments of IoT, increased connectivity and complexity of IoT ecosystems also introduce various vulnerabilities, making them attractive targets for attacks. 
Consequently, IoT security has become a critical issue for individuals, organizations, and governments worldwide.

Over the decades, IoT security has garnered significant attention from researchers, covering both defensive~\cite{kouicem2018internet,ahmad2021machine,williams2017identifying} and offensive~\cite{deogirikar2017security,gormucs2018security} strategies. 
As the large language model (LLM) has made significant strides in recent years, it has been explored in the context of IoT security as well, such as threat/vulnerability identification~\cite{sokiotllm,ma,llmiotfuz,yang2023iot}, perceive IoT sensor data~\cite{mo2024iot}, device management and labeling~\cite{meyuhas2024iotlabel,llmiotcontrol}, and IoT trust semantics enhancements~\cite{ferraris2024ici}.
This surge in interest has led to many domain-specific datasets that offer extensive insights covering various aspects of IoT security, such as vulnerabilities and exploits~\cite{CVE_Mission,VARIoT_db}, tactics, techniques, and procedures (TTPs)~\cite{strom2018mitre}, and industry-standard guidelines~\cite{abdul2019comprehensive,wright2022regulating}. 
However, most existing works focus on discovering new vulnerabilities/threats or designing novel defensive/offensive techniques, but pay less emphasis on leveraging the insights contained in datasets to assist or guide users in enhancing their security practices.
Although sources such as \cite{CVE_Mission,VARIoT_db,strom2018mitre} offer information search services, they typically only provide raw data (\eg, vulnerability descriptions) which is often challenging for non-technical users to understand. Even experienced security analysts may find it difficult to extract actionable insights from massive unprocessed information.
Consequently, \textit{there is an urgent need for solutions that deliver timely, actionable, and easily accessible IoT security and threat intelligence to a wide range of users, including both technical and non-technical ones.}

Inspired by LLM's advanced capability, there are two promising approaches \romannumeral1) \textit{fine-tuning} the LLM specifically on IoT security information and \romannumeral2) \textit{RAG}, which \textit{R}etrieves IoT security information to \textit{A}ugment the LLM's \textit{G}eneration. 
However, na\"{i}ve application of these methods are still facing one or several of the following challenges:

\noindent \textbf{Challenge-\ding{182}:} \textbf{IoT Security Evolves Rapidly.}
The field of IoT security is continuously developing, with new vulnerabilities, exploits, and security protocols emerging regularly. 
For example, in August 2024, the VARIoT vulnerability dataset~\cite{VARIoT_db} received 79 updates, and cybersecurity websites almost report new developments daily\footnote{\scriptsize \url{https://www.bleepingcomputer.com/}} \footnote{\scriptsize \url{https://www.darkreading.com/}}.
As a consequence, fine-tuning an LLM on static IoT data would quickly become outdated, failing to provide the latest threats or innovations in IoT security.

\noindent \textbf{Challenge-\ding{183}:} \textbf{Heterogeneous Dataset Formats.} 
IoT security datasets come in a variety of formats, such as structured data (\eg, vulnerability databases~\cite{VARIoT_db}), unstructured text reports, and product compliance lists (such as Cybersecurity Labelling Schemes in Singapore, Finland, Germany, and the United States).
Fine-tuning/augmenting an LLM on such diverse data types without a robust data processing mechanism would significantly limit the utilization of information contained in these datasets.

\noindent \textbf{Challenge-\ding{184}:} \textbf{Diverse User Requirements.} 
Users in IoT ecosystems range from consumers to security analysts, each with distinct expertise levels and specific security concerns. 
Na\"{i}vely combining LLM with IoT security information would struggle to effectively cater to these varied requirements, limiting its ability to provide meaningful and contextually appropriate responses for different user groups.

In light of the knowledge gap and technical challenges analyzed above, we ask the following question: 

\textit{Can we use large language models to effectively disseminate IoT security assistance to various key users of the IoT ecosystems in an understandable and actionable manner to provide better IoT security guarantees?}

We present \chatiot, an LLM-based IoT security assistant augmented with external information retrieval, to answer this question affirmatively. 
On the paradigm of RAG, we design \chatiot\ by combining the advanced language understanding and reasoning capabilities of LLM with fast-evolving IoT security-specific information to generate reliable, relevant, and technical answers.
To handle heterogeneous datasets, we have developed an end-to-end data processing toolkit that integrates a range of existing technologies, enabling the conversion of diverse data formats into retrievable documents and optimizing chunking strategies to improve retrieval performance.
Additionally, we define several use-case specifications and provide the user's \textit{background} to ensure that the LLM-generated answers are also user-friendly, \aka, aligned with each user's specific needs and expertise levels.
%Finally, we implement a prototype of \chatiot\ and perform extensive evaluations to assess the quality of the generated outputs.
In summary, \chatiot\ offers the following contributions:

\begin{itemize}
    \item \textbf{IoT security assistant driven by LLM \& threat intelligence.}
    Our IoT security assistant, \chatiot, is designed to leverage advanced LLM understanding and reasoning capabilities and threat intelligence at the same time. 
    We develop self-querying retrievers for diverse IoT threat datasets and dynamically activate retrievers based on the user's background and the query contexts. 
    In this way, LLM extracts IoT threat intelligence from the relevant retrieved documents.
    This integration allows the assistant to generate the latest reliable, relevant, and technical IoT security answers tailored to both query contexts and users' requirements.
    To our best knowledge, it is the first time to make use of LLM and IoT threat intelligence to offer IoT security assistance.

    \item \textbf{Data processing toolkit.}
    We design toolkit \datakit\ to handle datasets in diverse formats. Our toolkit first parses the raw data and converts the parsed contents into text. Notably, we utilize LLM to \romannumeral1) select the appropriate content for page\_content and metadata of documents, and \romannumeral2) process the multimodal contents, such as figures, tables, and codes, into text summarization. 
    Additionally, we integrate the Ragas library~\cite{es2023ragas} to optimize the chunking strategy, including splitter methods and chunk sizes. 
    While many of these technologies are adapted from existing works, our focus is on developing an end-to-end data processing toolkit, which might be useful and of independent interest.

    \item \textbf{Implementation \& Evaluation.}
    We implement a prototype of \chatiot\ and define five use cases. 
    For each case, we specify the user’s background in terms of \textit{knowledge}, \textit{goals}, and \textit{requirements} to guide the LLM in generating answers that are not only reliable, relevant, and technical but also user-friendly.
    Extensive evaluations show the improvements achieved by \chatiot. 
    Specifically, we compare \chatiot's answers with those generated directly from LLaMA3, LLaMA3.1, and GPT-4o in terms of reliability, relevance, technical depth, and user-friendliness\footnote{The metrics are explained in \S~\ref{sec:exp-llmeval}}.
    When evaluated with LLaMA3:70B, \chatiot\ improves all metrics by over $10\%$ on average, particularly in relevance and technicality. 
    Human evaluations also confirm that \chatiot\ provides better answers.
\end{itemize}

\smallskip
\noindent \textbf{Organization.}
We introduce the background in~\S~\ref{sec:back}. 
An overview of design goals and system architecture is given in~\S~\ref{sec:overgoal}.
Our concrete design is illustrated in~\S~\ref{sec:design}. 
We implement the prototype and perform experimental evaluations in~\S~\ref{sec:experiment}. 
Finally, we summarize the related works in~\S~\ref{sec:relatedwork} and conclude this work in~\S~\ref{sec:con}. 

\section{Background}\label{sec:back}

\subsection{IoT Threat Intelligence}\label{sec:iotintell}
IoT threat intelligence has emerged as a critical element in addressing security challenges, it refers to the collected and analyzed data related to threats, vulnerabilities, and attack patterns. 
In the context of IoT, threat intelligence helps to monitor and analyze threats specific to IoT ecosystems, including device vulnerabilities~\cite{VARIoT_db}, communication protocols~\cite{dragomir2016survey}, malware~\cite{alrawi2021circle}, tactics, techniques, and procedures (TTPs)~\cite{strom2018mitre} used by attackers, and others~\cite{iacovazzi2023towards,bou2020cyber,wagner2019cyber}.
Recent research has highlighted the importance of IoT threat intelligence in identifying zero-day vulnerabilities~\cite{saurabh2024hms} and mitigating threats using collaborative defense mechanisms~\cite{ahmed2023securing}.
The dynamic and fast-evolving nature of IoT systems, coupled with their diverse deployments, including both consumer and industrial settings, highlight the need for up-to-date and automated threat intelligence solutions capable of addressing its unique security challenges.

\begin{figure*}[ht]
    \centering
    \includegraphics[width=\linewidth]{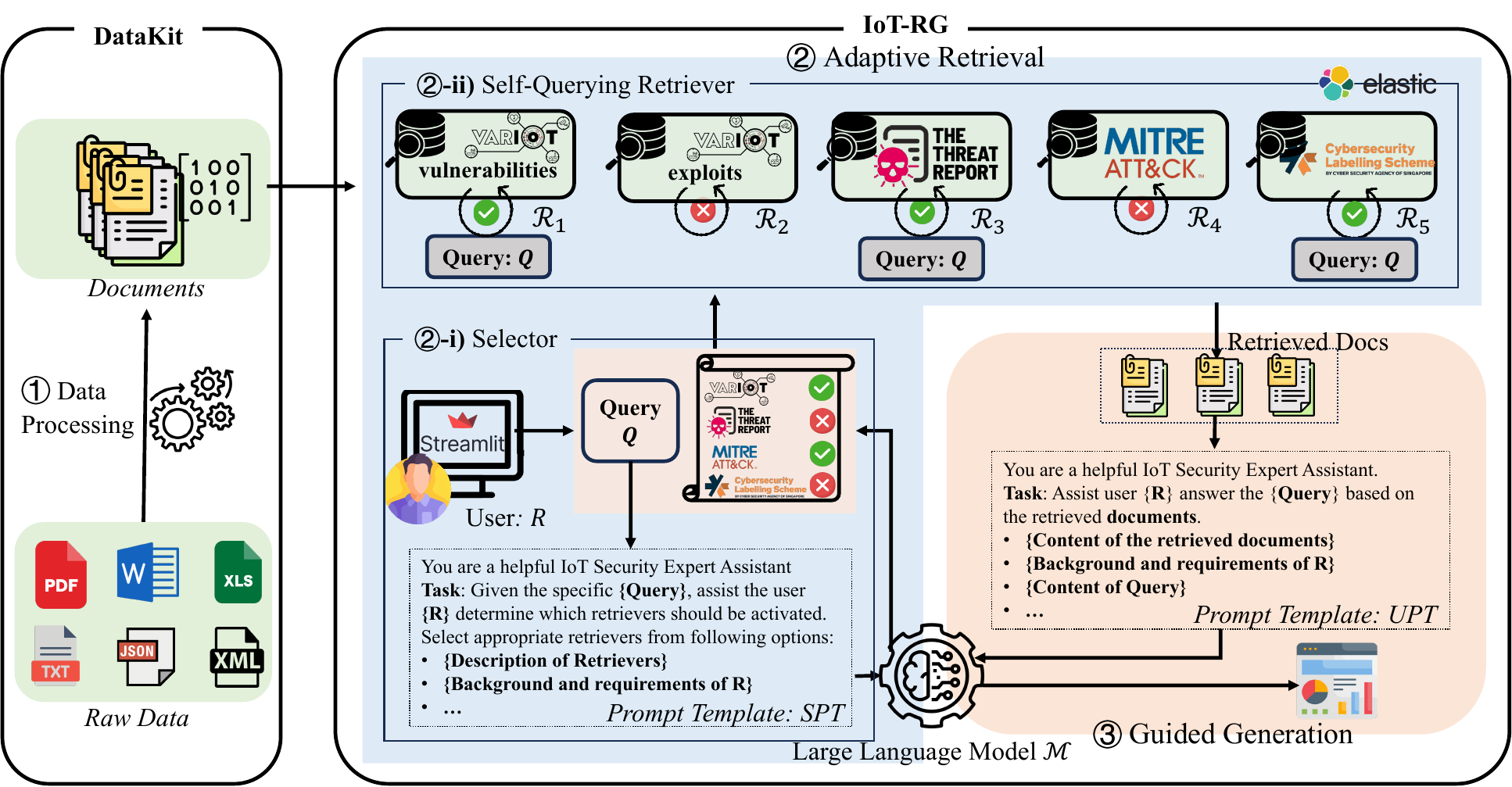}
    \caption{System architecture of \chatiot. Our system consists of modules \datakit\ and \iotrgen, and operates in three steps: \textcircled{1} \datakit\ processes IoT security datasets from various formats to documents for building retrievers. \textcircled{2} \iotrgen\ provides an interface for the user to submit the query and retrieves relevant documents adaptively: \romannumeral1) Selector first determines which retrievers should be activated, and \romannumeral2) The activated self-querying $\mathcal{R}_i$s retrieve similar documents and filter out unsatisfied ones based on metadata.
    \textcircled{3} LLM will synthesize all information to generate the answer.}
    \label{fig:system}
\end{figure*}

\subsection{Large Language Model}\label{sec:llm}
Large Language Model (LLM) is pre-trained on billions of available datasets, enabling it to capture vast amounts of linguistic, factual, and contextual knowledge. 
Due to the extensive training data, LLM can be directly employed for many downstream tasks, ranging from language translation to complex reasoning tasks~\cite{gpt,zhuge2021kaleido,docformer,kim2022ocr}.
However, LLM has certain limitations, particularly when it comes to handling highly specialized or constantly evolving domains.
Since LLM relies solely on the data they were trained on, which may not always reflect the latest information, it may generate incomplete or outdated responses for specific queries.
This is where Retrieval-Augmented Generation (RAG) comes into play.

RAG~\cite{fan2024survey,lewis2020retrieval} is a powerful approach that combines the generative capabilities of LLM with the latest external information retrieval. 
Instead of solely relying on the LLM's internal knowledge, RAG augments the LLM by retrieving relevant, up-to-date information from external databases or documents at the time of generation. 
The process involves querying an external knowledge source, retrieving the most relevant documents, and using the retrieved information to guide or enhance the LLM's response generation. 
By bridging the gap between static pre-trained knowledge and dynamic, real-world data, RAG enhances the performance of LLM on tasks that require both advanced language understanding, reasoning, and up-to-date information. This makes it particularly effective for specialized domains where access to the latest, domain-specific information is crucial.

\section{Overview}\label{sec:overgoal}
This section begins with our design goals (c.f., \S~\ref{sec:goal}) and then presents the system architecture about how to achieve the goals at a high level (c.f., \S~\ref{sec:overview}).

\subsection{Design Goals}\label{sec:goal}
Our goal in developing \chatiot\ is to equip it with the following essential features:

\begin{itemize}
\item\textbf{G-\textcircled{1}: Coping with the rapid evolution of IoT threat intelligence.}
\chatiot\ is designed to stay up-to-date with the rapidly evolving IoT threat intelligence.
By combining the advanced capabilities of LLM with external information retrieval, it is enhanced by the latest IoT threat information, such as new vulnerabilities, to deliver timely and technical insights.

\item \textbf{G-\textcircled{2}: Filtering relevant information.}
To prevent information overload, \chatiot\ uses advanced retrieval and filtering techniques to prioritize important, highly relevant information while discarding irrelevant data. This ensures that \chatiot\ generates actionable intelligence while avoiding overlooking critical information and filtering out unnecessary data.

\item \textbf{G-\textcircled{3}: Tailored for different user types.}
\chatiot\ guides its generated answers based on the users' roles (Consumer, Security Analyst, \etc) and their corresponding backgrounds.
This allows each kind of user to get insights or solutions that are relevant, understandable, and actionable based on their specific requirements and expertise level.
\end{itemize}

\subsection{System Architecture}\label{sec:overview}

As illustrated in Figure.~\ref{fig:system}, we designed \chatiot\ with two modules: the data processing toolkit, named \datakit, and the retrieval and generation system, abbreviated as \iotrgen, to achieve our goals:

\begin{itemize}
\item \textbf{Data Processing.} First, we integrate a variety of technologies to build our end-to-end data processing toolkit \datakit.
It can convert the collected IoT threat datasets, which are provided in various formats, into \textit{documents}, which are suitable for retrieval and LLM processing.
For each dataset, it extracts specific content for page\_content and metadata for documents, and a corresponding retriever is subsequently constructed.
\datakit\ ensures \chatiot\ can keep up with the rapid evolution of IoT threat intelligence (G-\textcircled{1}).

\item \textbf{Adaptive Retrieval.} Second, we achieve adaptive retrieval through \textit{Selector} and \textit{Self-Querying} mechanisms.
When a user submits a query $Q$, the Selector will invoke the equipped LLM to generate a configuration that determines which retrievers to activate. 
Once the configuration and query are passed to the retrievers, \iotrgen\ executes the activated self-querying retrievers to get highly relevant documents while filtering out irrelevant ones. 
This approach prevents overload, ensuring that only relevant and important documents will be utilized (G-\textcircled{2}).

\item \textbf{Guided Generation.} Finally, \iotrgen\ synthesizes the user’s background, $Q$, and retrieved documents to instruct the LLM to generate answers.
In this way, the answers are generated not only based on the advanced language understanding capabilities of LLM and the retrieved IoT security-related information, but also aligned with the user's background (G-\textcircled{3}).

\end{itemize}

In this way, \chatiot\ effectively provides reliable, relevant, and technical insights on IoT security tailored to different users.

\section{Design of \chatiot}\label{sec:design}
In this section, we first describe the construction of \iotrgen\ in \S~\ref{sec:retrieval}, followed by an overview of the data processing toolkit in \S~\ref{sec:toolkit}.
Finally, we present the use case specifications and explain how \chatiot\ is designed to be user-friendly in \S~\ref{sec:usecase}.

\subsection{Construction of \iotrgen}\label{sec:retrieval}

As shown in Figure.~\ref{fig:system}, \iotrgen\ consists of \textit{Adaptive Retrieval} and \textit{Guided Generation}.
We present their detailed constructions as follows.
%The third component utilizes LLM $\mathcal{M}$ to generate the response based on retrieved documents, role, and query, following a straightforward process. 
%This subsection will provide the detailed constructions of the Retriever Selector and Retrievers.

\subsubsection{Adaptive Retrieval}\label{sec:rca}
Recall that we have multiple retrievers, each dedicated to retrieving documents generated from a specific dataset. We achieve adaptive retrieval mechanism works in two aspects: 
\romannumeral1) selecting which retrievers should be activated and \romannumeral2) trying to retrieve only relevant documents while discarding irrelevant ones, even for the activated retrievers.

\smallskip
\noindent \textbf{Design of Selector.}
When user \textsf{role} submits a query $Q$, a straightforward and \textit{static} approach is to $Q$ to all retrievers and gather retrieved documents from them. 
However, this method has the following drawbacks:
\romannumeral1) \textit{Irrelevant Retrievers.} Documents of some retrievers may not be relevant to the query. Retrieving them not only fails to improve the quality of the generated answer but may even negatively affect it.
\romannumeral2) \textit{Resource Costs.} Retrieving unnecessary retrievers leads to increased resource costs, such as computational overhead, during both the retrieval and generation processes.

To address these issues, we introduce an LLM-based adaptive \textit{Selector}. As shown in Figure~\ref{fig:system}, we put the user \textsf{role} with the background, the query $Q$, and the descriptions of all retrievers as \textit{Selector Prompt (SPT)} and input SPT to LLM $\mathcal{M}$, which generates $\{\mathcal{S}_i\}_{i=1}^n$,
where $\mathcal{S}_i = \mathsf{True}$ implies that the retriever $\mathcal{R}_i$ should be activated, and $\mathsf{False}$ indicates not. 
In Table~\ref{tab:selectorexample}, we present the configurations generated by LLaMA3:8B for some example queries submitted by different users.

\begin{table}[ht]
    \centering
    \caption{Selector configurations $\{\mathcal{S}_i\}_{i=1}^n$ generated by LLaMA3:8B for different user roles and example queries. \ding{52} indicates \texttt{True} and \ding{56} is for \texttt{False}.}
    \label{tab:selectorexample}
    \resizebox{0.48\textwidth}{!}{
    \begin{tabular}{@{}p{1.5cm}|p{6cm}|c c c c c}
    \toprule
    \toprule
    \multicolumn{1}{c|}{\textbf{Role}} & \multicolumn{1}{c|}{\textbf{Example Query}} & $\mathcal{S}_1$ & $\mathcal{S}_2$ & $\mathcal{S}_3$ & $\mathcal{S}_4$ & $\mathcal{S}_5$  \\ \midrule
    Consumer & Is it secure to use Signify Smart Lighting in home? & \ding{52} & \ding{56} & \ding{52} & \ding{56} & \ding{52} \\ \midrule
    Security Analyst & Conduct a security assessment, including vulnerability, exploits, and threats, for TP-Link AX6000 Wi-Fi Router. & \ding{52} & \ding{52} & \ding{52} & \ding{56} & \ding{56} \\ \midrule
    Technical Officer & Check TTPs and security labeling of the company's WiFi Routers, including TP-Link, D-Link, and ASUS in Singapore. & \ding{52} & \ding{52} & \ding{56} & \ding{52} & \ding{52} \\ \midrule
    Developer & Develop a security enhancement roadmap for the next generation of TP-Link Wi-Fi routers. & \ding{52} & \ding{56} & \ding{52} & \ding{56} & \ding{52} \\ \midrule
    Trainer & Prepare a guide on the importance of cybersecurity labeling for smart locks like the August Smart Lock. & \ding{52} & \ding{56} & \ding{56} & \ding{56} & \ding{52}  \\
    \bottomrule
    \bottomrule
    \end{tabular}
    }
\end{table}

\begin{figure}[h]
    \centering
    \includegraphics[width=0.48\textwidth]{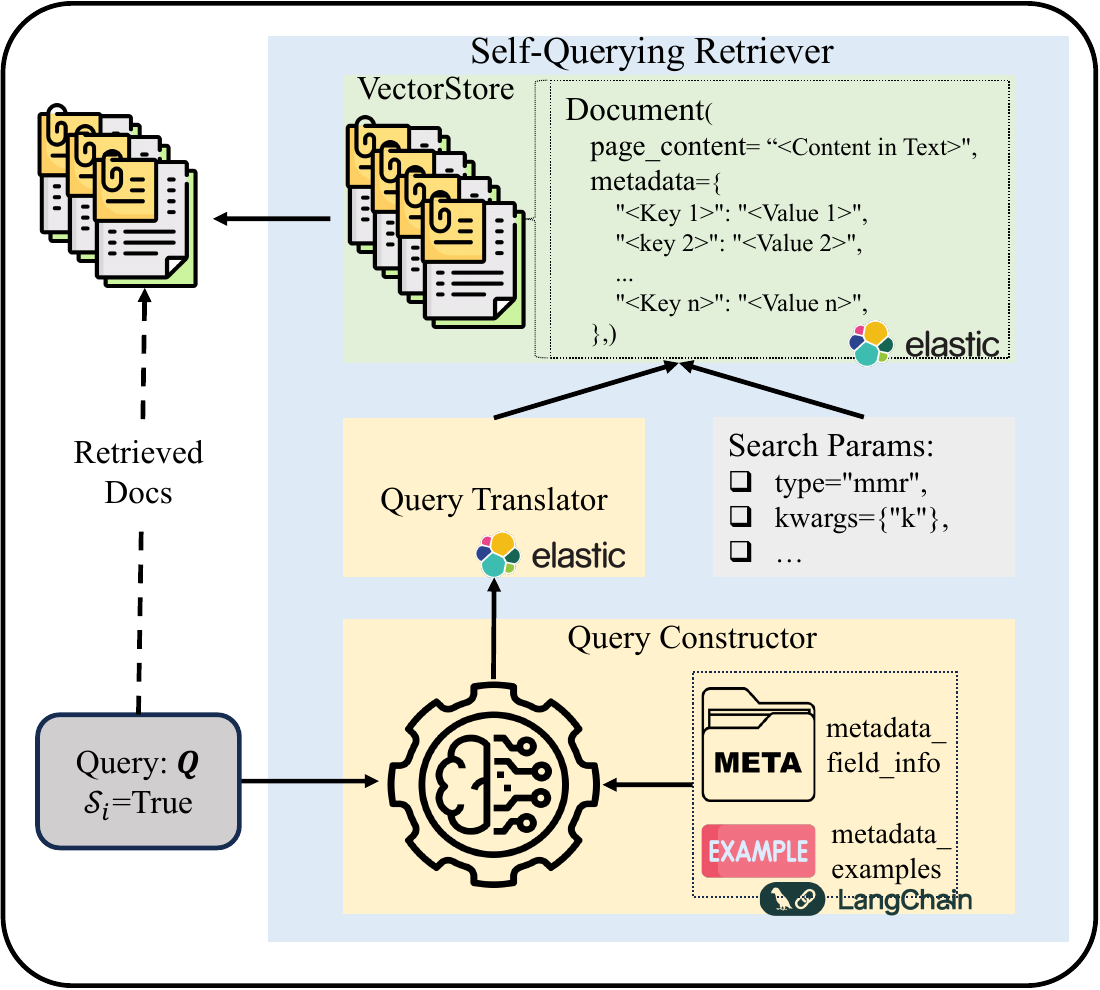}
    \caption{The construction of self-querying retriever based on LangChain and Elastic. When $\mathcal{S}_i=\mathsf{True}$, retrieve documents that are semantically similar to query $Q$ and filtered by metadata.}
    \label{fig:retri}
\end{figure}

\smallskip
\noindent \textbf{Self-Querying Retrievers.}
However, even the activated retrievers may return information that does not meet requirements, \eg, mismatch \textit{id} and \textit{products}. To address this problem, we make use of a self-querying technique to filter documents by \textsf{metadata}.

\begin{figure*}[h]
    \centering 
\includegraphics[width=\linewidth]{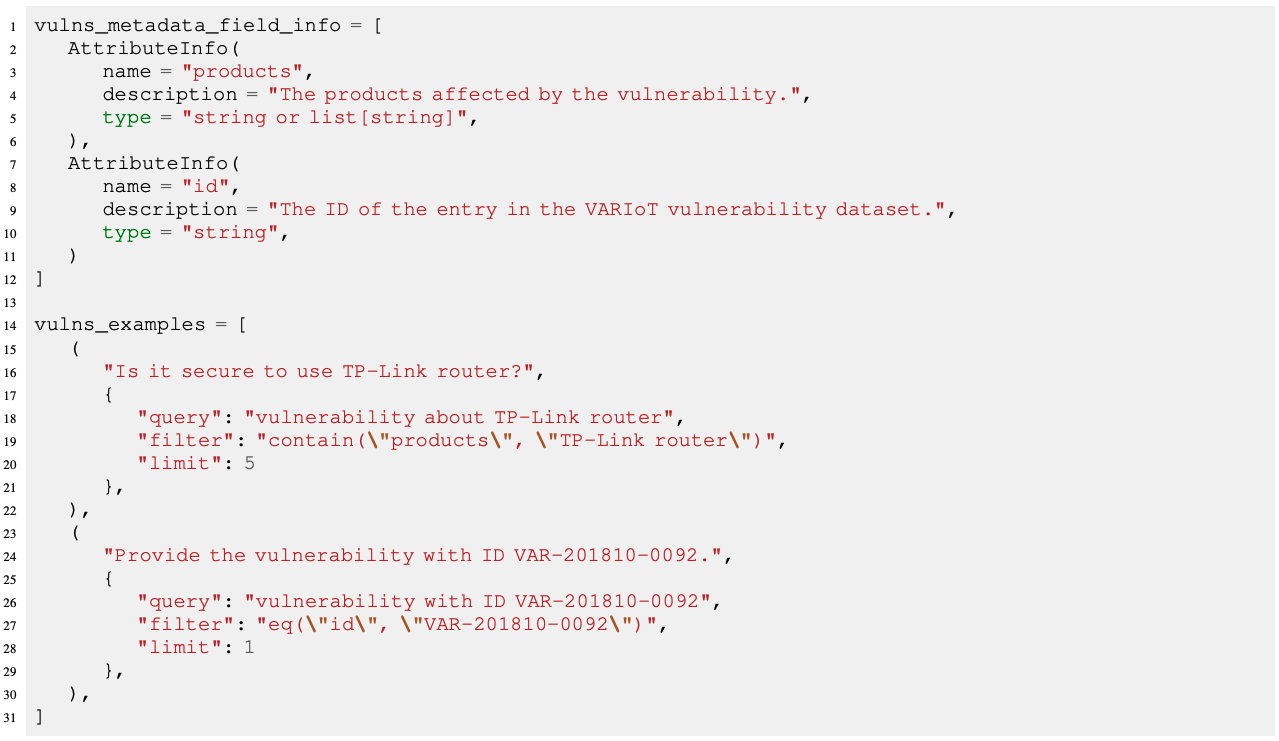}
\caption{Metadata fields information and examples for the self-querying retriever corresponding to the VARIoT vulnerabilities dataset.}
\label{fig:variotvulcode}
\end{figure*}

\begin{figure*}
    \centering
    \includegraphics[width=\linewidth]{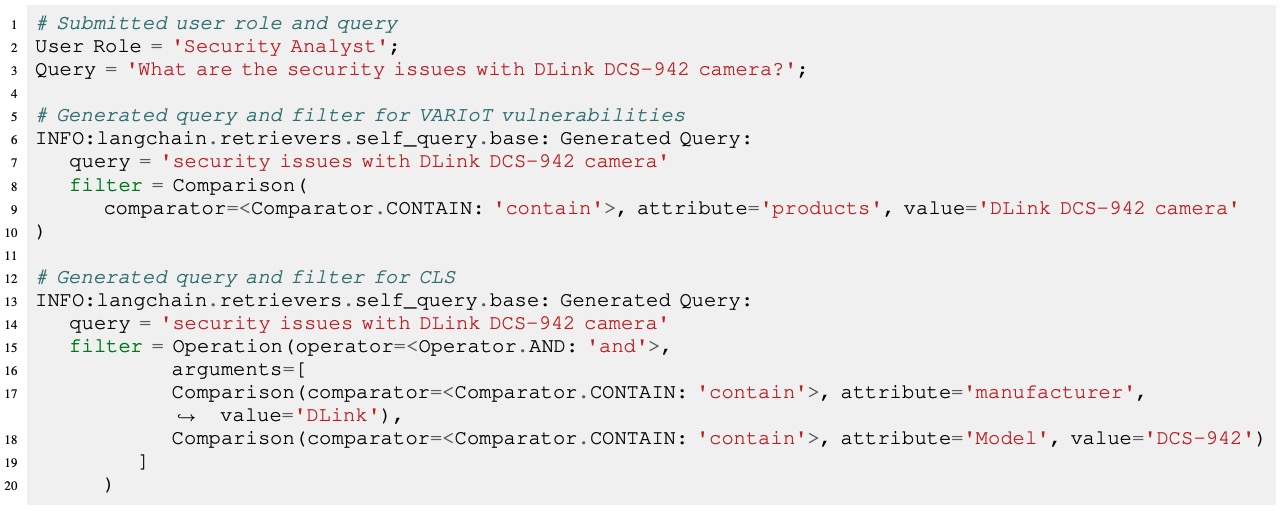}
    \caption{The generated queries and filters for VARIoT vulnerabilities and CLS list for the submitted query "What are the security issues with DLink DCS-942 camera?" by a Security Analyst.}
    \label{fig:querytrans}
\end{figure*}

As illustrated in Figure~\ref{fig:retri}, the self-querying retriever is composed of a Query Constructor, Query Translator, Search Params, and VectorStore.
When the user's query $Q$ is passed to Query Constructor, LLM will generate \textit{internal query language elements} based on pre-defined \textit{metadata\_field information} and \textit{metadata\_examples}.
Query Translator converts these elements into a \textit{structured query} with appropriate filters. 
Finally, the structured query and search parameters are applied to the Vector Store to retrieve documents. 
To specialize the self-querying retriever for IoT security, we take the following steps:
\begin{itemize}
    \item[\textcircled{1}] \textbf{Metadata \& Examples.} We provide the metadata field information and examples from relevant datasets. For instance, in the VARIoT vulnerabilities dataset, fields \textit{id} and \textit{products} are utilized as metadata. The corresponding \textit{metadata\_field\_info} and \textit{examples} are illustrated in Figure~\ref{fig:variotvulcode}, And details for other datasets are available in Appendix~\ref{appendix:metadataother}. This ensures the LLM can gain the necessary IoT security-specific knowledge to generate effective internal query language elements from the user's query.
    
    \item[\textcircled{2}] \textbf{Create Structured Queries.} Based on the above customized internal query language elements, Query Translator can create specific structured queries for each retriever. 
    Figure~\ref{fig:querytrans} shows how to enable the retrieval of VARIoT vulnerabilities and CLS lists that are both semantically similar to the query and appropriately filtered by their respective metadata.
\end{itemize}

%Specifically, given any query $Q$ in natural language, the retriever uses LLM-based to extract filters $F$ from $Q$, generate the structured query $\mathcal{Q}$ from $Q$ and $F$, and finally applies $\mathcal{Q}$ to its underlying VectorStore. 

\begin{figure}[h]
    \centering    
{\footnotesize
\begin{tikzpicture}
% Draw rounded rectangle with shadow
\node[rectangle, rounded corners, draw=black, fill=black!3!white, text width=0.43\textwidth, inner sep=12pt, align=left] (box) {
\textbf{\texttt{"Knowledge"}}: "The consumer may not have formal technical training but are familiar with using IoT devices for daily convenience such as smart home systems. The consumer has a basic understanding of device operation but may not be aware of the intricate security risks that exist."
\vspace{5pt}\\
\textbf{\texttt{"Goals"}}: "The primary aim is to understand whether a device is secure and how to maintain or improve its security, ensure safety, security, and reliability
of IoT devices within their homes or personal environments."
\vspace{5pt}\\
\textbf{\texttt{"Requirements"}}: "The answers should be practical, easy to follow, and focused on actionable steps the general user can take."
};
\end{tikzpicture}
}
\caption{The background for consumer utilized to guide the \chatiot\ to generate consumer-friendly outputs.}
\label{fig:usecasedef-consumer}
\end{figure}

\subsubsection{Guided Generation}\label{sec:resgen}

After retrieval, one direct step is feeding the retrieved documents and query to LLM to generate the answer. 
However, this simple approach is likely to result in user-unfriendly outputs. 
For example, consumers often lack the expertise needed to fully comprehend highly technical content, making such answers unhelpful and not actionable for them.

To address this issue, we incorporate user-specific backgrounds, including knowledge, goals, and requirements, into each user type's user-friendly prompt template \textit{UPT}. This adjustment guides LLM in generating answers tailored to the user's needs.
The specific background for the general consumer is shown in Figure~\ref{fig:usecasedef-consumer}, demonstrating how we simplify content for easy understanding. The background specifications of other user types can be referred to Appendix~\ref{appendix:bk}.

\begin{algorithm}[t]
\caption{\iotrgen}\label{alg:iotrgen}
\begin{algorithmic}[1]
\REQUIRE
User \textsf{role}, query $Q$, large language model $\mathcal{M}$, and retrievers $\{\mathcal{R}_i\}_{i=1}^n$.

\ENSURE
Generated answer $A$.

\STATE
\textcolor{gray}{$\triangleright$ \textbf{Procedure of Selector:}}

\STATE
Set prompt $\mathsf{SPT} = (\mathsf{Task}, \mathsf{role}, Q, \{\mathcal{R}_i\}_{i=1}^n)$, where $\mathsf{role}$ includes user's background implicitly and $\mathcal{R}_i$ indicates its description here.
\STATE
Inputting $\mathsf{SPT}$ to $\mathcal{M}$ and get $\{S_i\}_{i=1}^n = \mathcal{M}(\mathsf{SPT})$, where $S_i\in \{\mathsf{True}, \mathsf{False}\}$.

\STATE 
\textcolor{gray}{$\triangleright$ \textbf{Procedure of Self-Querying Retrieval:}}
\FORALL{$i=1, \dots, n$}
\IF{$\mathcal{S}_i = \mathsf{True}$}
\STATE
Execute self-querying retriever $\mathcal{R}_i$ and get documents $D_i=\mathcal{R}_i(Q)$
\ELSE
\STATE
Set $D_i = \mathsf{NULL}$.
\ENDIF
\ENDFOR

\STATE
\textcolor{gray}{$\triangleright$ \textbf{Procedure of Guided Generation:}}

\STATE
Set prompt $\mathsf{UPT} = (\mathsf{Task}, \mathsf{role}, Q, \{\mathcal{D}_i\}_{i=1}^n)$

\RETURN
$A = \mathcal{M}(\mathsf{UPT})$.

\end{algorithmic}
\end{algorithm}

Formally, we summarize and show the workflow of \iotrgen\ in algorithm~\ref{alg:iotrgen}.

\begin{figure*}
    \centering
    \includegraphics[width=\linewidth]{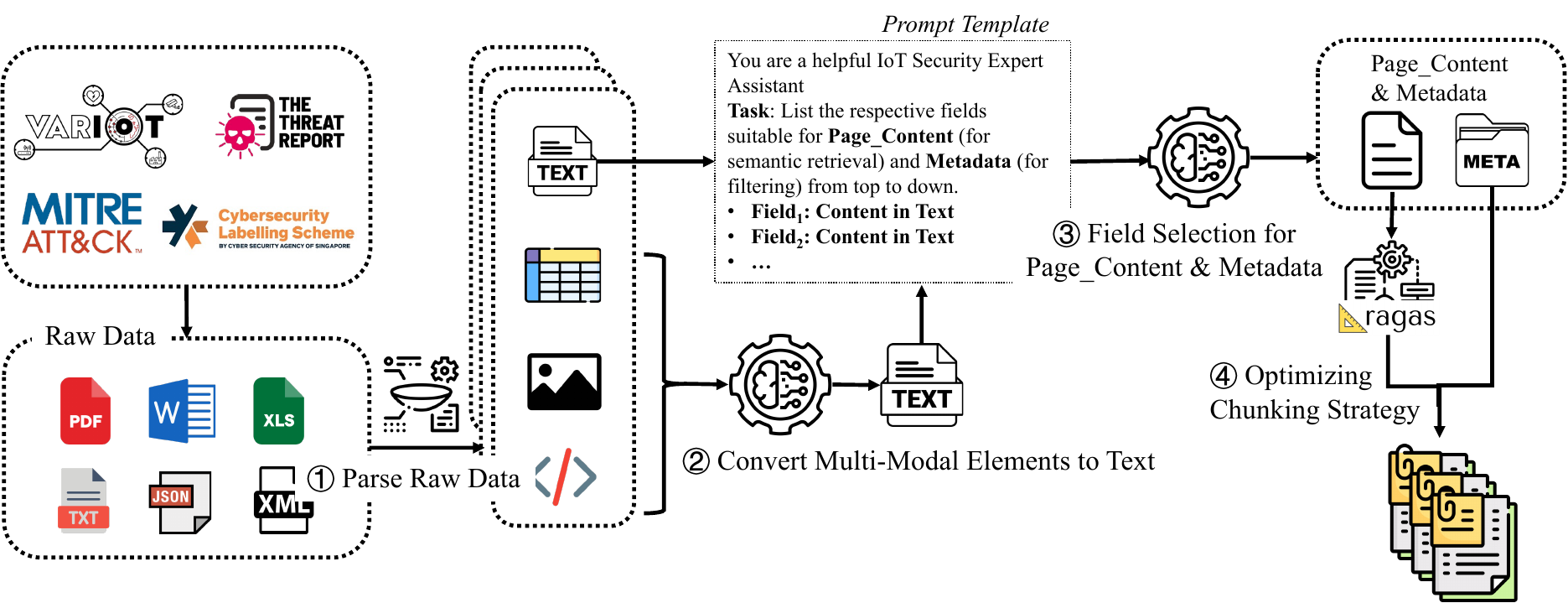}
    \caption{The construction of data processing toolkit \datakit. After collecting various IoT security datasets, \datakit\ first parses the raw data to get elements of multi-modal (step~\textcircled{1}), and then converts the multi-modal elements into text by utilizing LLM (step~\textcircled{2}). Finally, \datakit\ uses LLM to select fields for the page\_content and metadata of documents (step~\textcircled{3}), and optimizes the chunking strategy (step~\textcircled{4}).}
    \label{fig:dataprocess}
\end{figure*}

\subsection{Data Processing Toolkit}\label{sec:toolkit}

The raw data of the collected datasets are of different formats, \eg, PDF and JSON. These formats are not suitable for RAG processing directly, and thus we develop our end-to-end data processing \datakit. As shown in Figure~\ref{fig:dataprocess}, \datakit\ works in four steps:
\begin{itemize}
    \item[\ding{172}] \textbf{Parse Raw Data.} The initial step involves parsing raw data into distinct content elements such as text, tables, figures, and code. Leveraging existing tools like the unstructured library\footnote{\scriptsize \url{https://pypi.org/project/unstructured/}} helps in extracting textual content from threat reports, while the JSON library\footnote{\scriptsize \url{https://python.readthedocs.io/en/v2.7.2/library/json.html}} is useful for handling structured data from sources like VARIoT and MITRE ATT\&CK.
    
    \item[\ding{173}] \textbf{Convert Multi-Modal Elements to Text.} Once parsed, any multi-modal elements (\eg, tables, figures, code) must be converted into text descriptions for further processing. LLMs like LLaVA (for images)~\cite{liu2023llava,liu2023improvedllava,liu2024llavanext}, LLaMA3:8B (for tables), and CodeLlama (for code)~\cite{roziere2023code} are employed to generate these descriptions. This modular approach allows for easy integration of other LLMs to handle different types of multi-modal content.

    \item[\ding{174}] \textbf{Field Selection for Page\_Content \& Metadata.} In structured formats like JSON, content is often stored across multiple fields.
    Instead of using all fields, it is crucial to identify and utilize the most relevant ones for retrievers. This is done by sampling example items from each field and using an LLM (\eg, LLaMA3:8B) to intelligently select fields that best represent the document's page\_content and metadata, and the prompt is shown in Figure~\ref{fig:dataprocess}.
    Concretely, the selected fields for page\_content and metadata of each dataset are shown in \S~\ref{sec:exp-field}.

    \item[\ding{175}] \textbf{Optimize Chunking Strategy.} The final step involves selecting an appropriate chunking strategy, including the chunking size, overlap, and splitting method. The Ragas library~\cite{es2023ragas} is used to optimize this process, and the details are shown in algorithm~\ref{alg:chunkopt}.
\end{itemize}

\begin{algorithm}[t]
\caption{Optimize Chunking Strategy}\label{alg:chunkopt}
\begin{algorithmic}[1]
\REQUIRE
Documents $D$, chunking $sizes$, $overlaps$, and $splitters$ functions, $metrics=[precision, recall]$, and LLM $\mathcal{M}$.

\ENSURE
Optimized $(size^*, overlap^*, splitter^*)$.

\FORALL{$splitter \in splitters$}
\FORALL{$size \in chunk\_sizes$}
\FORALL{$overlap\in overlaps$}
\IF{$overlap < size$}
\STATE
\textcolor{gray}{$\triangleright$ \textbf{Split $D$ as small chunks:}}
\STATE
$\{d_i\}_i = splitter(D, size, overlap)$.
\STATE
$(p, r) = \text{Ragas.evaluate}(\{d_i\}_i, \mathcal{M}, metrics)$.
\ELSE
\STATE
\texttt{break.} \textcolor{gray}{$\triangleright$ \textbf{Go to next chunk size.}}
\ENDIF
\ENDFOR
\ENDFOR
\ENDFOR
\RETURN
chunking $(size^*, overlap^*, splitter^*)$ with a optimized trade-off between $(p, r)$.
\end{algorithmic}
\end{algorithm}

After obtaining the optimized chunking strategy, we split the documents into small chunks and use the all-MiniLM model for chunked text embedding. 
The documents are composed of chunked text, embedding, and metadata.
This approach ensures that the IoT security and threat datasets are processed efficiently and ready for further analysis or use in LLM. 
Note while many of these technologies are adapted from existing works, we foucs on putting them all together to develop an end-to-end data processing toolkit, which might be of independent interest and useful in practical applications.

\subsection{Use Cases Specialization}\label{sec:usecase}

\begin{figure}
    \centering    
{\footnotesize
\begin{tikzpicture}
% Draw rounded rectangle with shadow
\node[rectangle, rounded corners, draw=black, fill=black!3!white, text width=0.43\textwidth, inner sep=12pt, align=left] (box) {
    \textbf{\texttt{"User Role"}}: \{\\
    \hspace{15pt} \textbf{"description"}: \textit{"The role of the user, such as Consumer."}, \\
    \hspace{15pt} \textbf{"type"}: \textit{"enum string"} \\
    \} \vspace{5pt}\\
    \textbf{\texttt{"Background"}}: \{\\
    \hspace{15pt} \textbf{"description"}: \textit{"Specifies the knowledge, goals, and requirements of user."}, \\
    \hspace{15pt} \textbf{"type"}: \textit{"enum string"} \\
    \} \\
    \textbf{\texttt{"Actions"}}: \{\\
    \hspace{15pt} \textbf{"description"}: \textit{"Lists the tasks the user can perform."}, \\
    \hspace{15pt} \textbf{"type"}: \textit{"list of strings"} \\
    \} \vspace{5pt}\\
    \textbf{\texttt{"Example Query"}}: \{\\
    \hspace{15pt} \textbf{"description"}: \textit{"Provides a sample question/query."}, \\
    \hspace{15pt} \textbf{"type"}: \textit{"string"} \\
    \}
};

\end{tikzpicture}
}
\caption{Use case specializations. For each use case, we define its role, goal, actions, and example query. For each property, we present its description and type.}
\label{fig:usecasedef}
\end{figure}

\begin{table*}[]
\centering
\caption{The actions and example query for each kind of user role defined for \chatiot.}
\label{tab:usecase}
\resizebox{\textwidth}{!}{

\begin{tabular}{p{1.5cm}|p{12cm}|p{4cm}}
%\begin{tabular}{@{}c@{\hskip 0.1cm}|
%    @{\hskip 0.1cm}>{\arraybackslash}p{10.5cm}@{\hskip 0.1cm}|
%    @{\hskip 0.1cm}>{\arraybackslash}p{5cm}@{\hskip 0.1cm}
%    }
\toprule
\toprule
\multicolumn{1}{c|}{Role} & \multicolumn{1}{c|}{Actions} & \multicolumn{1}{c}{Example Query}  \\ \midrule
Consumer  
& \romannumeral1) Assess the security of IoT devices before purchase or installation; \newline
\romannumeral2) Monitor ongoing security status and updates for existing devices;\newline
\romannumeral3) Make informed decisions based on security reports provided by \chatiot.
& Is it secure to use Signify Smart Lighting in home? \\
\midrule

Security Analyst  
&
\romannumeral1) Identify and evaluate security threats and vulnerabilities in IoT devices;\newline
\romannumeral2) Recommend mitigation strategies based on threat intelligence and analysis;\newline
\romannumeral3) Provide detailed security reports to stakeholders.
& Conduct a security assessment for TP-Link AX6000 Wi-Fi 6 Router. 
\\ \midrule

Technical Officer 
&
\romannumeral1) Ensure that IoT devices are deployed securely and operate within compliance guidelines;\newline
\romannumeral2) Oversee the application of security updates and patches;\newline
\romannumeral3) Monitor the security posture of the IoT ecosystem within their organization.
& Check the security labeling of the company's WiFi Routers,  including TP-Link, D-Link, and ASUS in Singapore. 
\\
\midrule

Developer
&
\romannumeral1) Design and develop secure IoT products by adhering to best practices and security standards;\newline
\romannumeral2) Continuously update products to address new vulnerabilities and threats;\newline
\romannumeral3) Provide accurate security documentation and updates to customers.
&
Develop a security enhancement roadmap for the next generation of TP-Link Wi-Fi routers.
\\
\midrule

Trainer 
&
\romannumeral1) Develop and deliver training programs on IoT security;\newline
\romannumeral2) Guide users and organizations on how to secure IoT devices and respond to incidents;\newline
\romannumeral3) Provide up-to-date information on IoT security trends and best practices.
&
Prepare a guide on the importance of cybersecurity labeling for smart locks like the August Smart Lock.
\\
\bottomrule
\bottomrule

\end{tabular}}
\end{table*}

In this section, we define five specialized use cases of \chatiot. 
Each use case is defined by \textit{four} fundamental properties: \textit{User Role}, \textit{Background}, \textit{Actions}, and \textit{Example Query}, within the IoT security domain.
The detailed specifications are illustrated in Figure~\ref{fig:usecasedef}.
The roles include \textit{Consumer}, \textit{Security Analyst}, \textit{Technical Officer}, \textit{Developer}, and \textit{Trainer}. 
Recall that we have discussed background in \S~\ref{sec:resgen}, we present the detailed actions and example query in the following context.

Table~\ref{tab:usecase} highlights the key actions associated with each user role, such as assessing the security of IoT devices, deploying security patches, or developing training programs on IoT security.
Additionally, it provides example queries for each role, demonstrating how \chatiot\ can be utilized to address the unique needs of various users. 
This structured approach ensures that \chatiot\ caters to a diverse range of users, offering tailored assistance and enhancing IoT security management across different scenarios.
Note that while we provide five use cases, they are not rigid or fixed. 
The use cases can be easily extended by defining new user roles, specifying the background (including knowledge, goals, and requirements), and outlining actions. Example queries can also be added to further clarify the context and functionality of each user role.

\section{Experimental Evaluation}\label{sec:experiment}

In this section, we implement \chatiot\ and conduct extensive evaluation. 
We study the effectiveness of \chatiot\ and answer the following questions:

\noindent \textbf{Q1:} How does \datakit\ extract appropriate field selection from IoT threat datasets and convert them into well-structured documents for retrieval and LLM analysis?
What are the optimal chunking strategies for each kind of document? (\S~\ref{sec:exp-dataprocess})

\noindent \textbf{Q2:} What are the advantages of our system? Can \chatiot\ effectively generalize and improve the capabilities of the most advanced LLMs available in processing IoT security issues? (\S~\ref{sec:exp-llmeval})

\noindent \textbf{Q3:} Can \chatiot\ be a practical useful IoT assistant over using LLM alone? How about the feedback from real-world human evaluation? (\S~\ref{sec:exp-human})

\subsection{Setup}
\noindent \textbf{Testbed.}
We implement \chatiot\ in Python 3.10.13, utilizing large language models LLaMA3:8B \& 70B and LLaMA3.1:8B \& 70B provided by Groq\footnote{\scriptsize \url{https://chat.groq.com/}}, GPT-4o-mini and 4o provided by OpenAI\footnote{\scriptsize \url{https://platform.openai.com/docs/models/gpt-4o}}. All these LLMs are utilized by calling their APIs.
For building the vector store, we employed Elasticsearch 8.13.2~\cite{elasticsearch2018elasticsearch}, running on Docker Desktop 4.29.0~\cite{docker-desktop}. 
All components were integrated using the LangChain library (version 0.2.5)~\cite{langchain2024}. 
The WebApp was developed using Streamlit (version 1.33.0)~\cite{streamlit}. 
Experiments were conducted on a MacBook Pro equipped with an Apple M3 Pro CPU (11 cores) and 18 GB of RAM, running macOS 14.6.1 with the Darwin 23.6.0 kernel.

\smallskip
\noindent \textbf{Data Sources.}
We collect five kinds of IoT security and threat datasets from the public Internet:
\begin{itemize}
\item[\romannumeral1)] VARIoT vulnerabilities~\cite{VARIoT_db}: This dataset catalogs known vulnerabilities in various IoT devices, offering detailed information about the potential risks associated with each vulnerability.
\item[\romannumeral2)] VARIoT exploits~\cite{VARIoT_db}: This dataset contains exploits targeting IoT devices, providing insights into the techniques and methods attackers use to compromise these systems.
\item[\romannumeral3)] MITRE ATT\&CK ICS TTPs~\cite{strom2018mitre}: This dataset outlines the tactics, techniques, and procedures (TTPs) employed by adversaries specifically in industrial control systems (ICS), which often include IoT-related TTPs as well.
\item[\romannumeral4)] Threat reports: We collect $17$ public threat reports from VXUG\footnote{\scriptsize \url{https://vx-underground.org/}} about emerging threats and vulnerabilities, offering analysis and recommendations for mitigating potential risks.
\item[\romannumeral5)] Cybersecurity labelling schemes\footnote{\scriptsize \url{https://www.csa.gov.sg/our-programmes/certification-and-labelling-schemes}}\footnote{\scriptsize \url{https://tietoturvamerkki.fi/en/products}}
\footnote{\scriptsize \url{https://www.nemko.com}}
\footnote{\scriptsize \url{https://nvlpubs.nist.gov/nistpubs/CSWP/NIST.CSWP.02042022-2.pdf}}: These schemes provide information on the security posture of various IoT products, helping consumers and organizations assess the security standards and certifications achieved by specific devices.
\end{itemize}
These datasets provide comprehensive insights into the current landscape of IoT threats, enabling us to enhance our system's capabilities.

%For each user case defined in \S~\ref{sec:usecase}, we propose 10 queries. 
%Then, we compare the generated answers of \chatiot\ with those generated by only using the corresponding LLM without augmented retrieval.
%and the commercial LLMs such as OpenAI ChatGPT3.5 and Google Gemini, to illustrate our improvements in IoT security.

\subsection{Fields Selection \& Chunking Strategy}\label{sec:exp-dataprocess}

This section shows the experimental evaluation for field selection and chunking strategy optimization.

\begin{table}[h]
    \centering
    \caption{Large Language Models-based field selection for page\_content and metadata. \ding{52} is for page\_content, $\bigcirc$ denotes metadata, and \ding{56} indicates unused fields.}
    \label{tab:fieldsel-eva}
    \resizebox{0.48\textwidth}{!}{
    \begin{tabular}{c|c|ccc}
    \toprule
    \toprule
    {Dataset} & {Fields} & {\footnotesize LLaMA3:8B} & {\footnotesize LLaMA3.1:70B} & {\footnotesize GPT-4o} \\ \midrule
    \multirow{12}{*}{\rotatebox{90}{VARIoT Vulnerabilities}} 
    & \texttt{cve} & \ding{56}  &$\bigcirc$ &  $\bigcirc$\\ 
    &\texttt{id} & $\bigcirc$ &$\bigcirc$ & $\bigcirc$\\ 
    &\texttt{credit} & \ding{52} & \ding{56} & $\bigcirc$\\ 
    &\texttt{description} & \ding{52} & \ding{52} & \ding{52} \\ 
    &\texttt{title} & \ding{52} & \ding{52} & \ding{52} \\ 
    &\texttt{products} & $\bigcirc$ &$\bigcirc$ & $\bigcirc$ \\ 
    &\texttt{vulns.-config.} & $\bigcirc$ & \ding{52} & $\bigcirc$ \\ 
    &\texttt{cvss-score} & \ding{56} & \ding{56} & $\bigcirc$ \\ 
    &\texttt{cvss-string} & \ding{56} & $\bigcirc$ & \ding{52} \\ 
    &\texttt{reference} & \ding{56} & \ding{52} & \ding{52} \\ 
    &\texttt{published} & \ding{56} & $\bigcirc$ & \ding{56} \\ 
    &\texttt{modified} & \ding{56} & $\bigcirc$ & \ding{56}\\ 
    \midrule
    \multirow{8}{*}{\rotatebox{90}{VARIoT Exploit}} 
    & \texttt{id} & $\bigcirc$ & $\bigcirc$ & $\bigcirc$ \\ 
    &\texttt{credit} & \ding{56} & $\bigcirc$ & $\bigcirc$ \\
    &\texttt{description} & \ding{52} & \ding{52} & \ding{52}\\
    &\texttt{exploit} & \ding{52} & \ding{52} & \ding{52}\\
    &\texttt{title} & \ding{52} & \ding{52} & \ding{52} \\
    &\texttt{cve-id} & $\bigcirc$ & $\bigcirc$ & $\bigcirc$ \\
    &\texttt{reference} & \ding{56} & $\bigcirc$ & \ding{52} \\
    &\texttt{products} & $\bigcirc$ & $\bigcirc$ & $\bigcirc$ \\
    \midrule
    \multirow{12}{*}{\rotatebox{90}{MITRE ATT\&CK ICS}} 
    &\texttt{stixId} & $\bigcirc$ & $\bigcirc$ & $\bigcirc$ \\
    &\texttt{name} & \ding{52} & \ding{52} & \ding{52} \\
    &\texttt{parentName} & \ding{52} & \ding{56} & \ding{52} \\
    &\texttt{description} & \ding{52} & \ding{52} & \ding{52} \\
    &\texttt{lunrRef} & \ding{56} & \ding{56} & $\bigcirc$ \\
    &\texttt{id} & $\bigcirc$ & \ding{56} & $\bigcirc$ \\
    &\texttt{url} & \ding{56} & \ding{56} & \ding{56} \\
    &\texttt{is\_enterprise} & \ding{56} & $\bigcirc$ & $\bigcirc$ \\
    &\texttt{type} & \ding{56} & $\bigcirc$ & $\bigcirc$ \\
    &\texttt{relatedTech.} & \ding{56} & $\bigcirc$ & \ding{56} \\
    &\texttt{is\_ics} & \ding{56} & \ding{56} & \ding{56} \\
    &\texttt{deprecated} & \ding{56} & \ding{56} & $\bigcirc$ \\
    \bottomrule
    \bottomrule
\end{tabular}}
\end{table}

\subsubsection{Fields for Page\_Content \& Metadata}\label{sec:exp-field}
Recall that for each dataset, we should first determine the fields for page\_content and metadata for documents before building self-querying retrievers (see \S~\ref{sec:toolkit}).
For each dataset, we sample 3 items, list the fields' names, and instruct the LLM to select the suitable fields. The results are as follows:
\begin{itemize}
    \item We use three LLMs: LLaMA3:8B, LLaMA3.1:70B, and GPT4-o, to select fields for VARIoT vulnerabilities, exploits, and MITRE ATT\&CK ICS.
    The experimental results are summarized in Table~\ref{tab:fieldsel-eva}.
    While there are slight variations in their selections, there is consensus on the crucial decisions:
    For instance, in the case of the VARIoT vulnerabilities, all LLMs select \texttt{title} and \texttt{description} for page\_content, and \texttt{id} and \texttt{products} for metadata.
    Similar selections are observed for the VARIoT exploits and MITRE ATT\&CK ICS datasets.

    \item For threat reports, which are typically unstructured, we use the report's content as page\_content and the title as metadata (Note we do not use self-querying retrieval for threat reports). 
    The CLS schemes consist solely of metadata with no descriptive content, so we leave page\_content blank and utilize the metadata for self-querying retrievers.
\end{itemize}
The specific fields selected for page\_content and metadata are detailed in Table~\ref{tab:fieldselect}.

\begin{table}[]
    \centering
    \caption{Selected fields for page\_content and metadata of each dataset used in \chatiot.}
    \label{tab:fieldselect}
    \resizebox{0.48\textwidth}{!}{
    \begin{tabular}{@{}c@{\hskip 0.1cm}|@{\hskip 0.1cm}>{\centering\arraybackslash}p{4cm}@{\hskip 0.1cm}|@{\hskip 0.1cm}>{\centering\arraybackslash}p{2cm}@{\hskip 0.1cm}}
    \toprule \toprule
         {Dataset} & {Page\_content} & {Metadata}  \\ \midrule
         {\scriptsize {VARIoT Vulns.}} & {\scriptsize {\texttt{title}, \texttt{description}}} & {\scriptsize {\texttt{id}, \texttt{products}}}\\ \hline 
         {\scriptsize{VARIoT Exps.}} & {\scriptsize {\texttt{title}, \texttt{description}, \texttt{exploit}}} & {\scriptsize {\texttt{id}, \texttt{products}}} \\ \hline
         {\scriptsize{ICS}} & {\scriptsize {\texttt{name}, \texttt{description}}} & {\scriptsize {\texttt{stixId}}} \\ \hline
         {\scriptsize{Threat Report}} & {\scriptsize {\texttt{Report's content}}} & {\scriptsize \texttt{title}} \\ \hline
         {\scriptsize{CLS}} & {\scriptsize \texttt{NULL}} & {\scriptsize {\texttt{All Fields}}} \\
         \bottomrule \bottomrule
    \end{tabular}}
\end{table}

\subsubsection{Chunking Evaluation}\label{sec:exp-chunk}

\begin{table*}[]
\centering
\caption{Context (precision, recall) of different chunking configurations for VARIoT Vulnerabilities, Exploits, MITRE ATT\&CK ICS, and Threat Reports. The best (precision, recall) in our experimental settings are marked in bold.}
\label{table:chunking}
\resizebox{\textwidth}{!}{
\begin{tabular}{c|ccc|ccc}
\toprule
\toprule
\multirow{2}{*}{(Size, Overlap)} &\multicolumn{3}{c|}{VARIoT Vulnerabilities} & \multicolumn{3}{c}{VARIoT Exploits} \\
& \texttt{Character} & \texttt{RecurChar} & \texttt{TokenText} & \texttt{Character} & \texttt{RecurChar} & \texttt{TokenText} \\
\midrule
(500,50) & (0.875, 0.906) & (0.981, 0.898) & (0.936, 0.896)
& (0.937, 0.903) & (0.927, 0.922) & (0.927, 0.896)\\
(500,100) & (0.876, 0.921) & \textbf{(0.986, 0.973)} & (0.902, 0.925)
& (0.918, 0.902) & (0.883, 0.908) & (0.910, 0.935)\\
(500,150) & (0.917, 0.909) & (0.983, 0.955) & (0.929, 0.917)
& (0.918, 0.895) & (0.912, 0.823) & (0.924, 0.884)\\
(500,200) & (0.929, 0.891) & (0.913, 0.935) & (0.937, 0.897)
& (0.927, 0.913) & (0.913, 0.867) & (0.901, 0.955)\\
\midrule
(1000,50) & (0.905, 0.867) & (0.955, 0.892) & (0.915, 0.902)
& (0.799, 0.898) & (0.871, 0.811) & (0.937, 0.920)\\
(1000,100) & (0.883, 0.883) & (0.923, 0.897) & (0.888, 0.843)
& (0.794, 0.908) & (0.878, 0.753) & (0.941, 0.920)\\
(1000,150) & (0.891, 0.897) & (0.916, 0.959) & (0.892, 0.873)
& (0.908, 0.751) & (0.920, 0.722) & \textbf{(0.943, 0.941)}\\
(1000,200) & (0.891, 0.918) & (0.913, 0.921) & (0.899, 0.883)
& (0.892, 0.846) & (0.804, 0.864) & (0.935, 0.925)\\
\midrule
(1500,50) & (0.895, 0.882) & (0.893, 0.922) & (0.918, 0.829)
& (0.851, 0.920) & (0.922, 0.861) & (0.917, 0.962)\\
(1500,100) & (0.828, 0.881) & (0.891, 0.913) & (0.901, 0.831)
& (0.872, 0.862) & (0.898, 0.904) & (0.934, 0.923)\\
\midrule
(1500,150) & (0.894, 0.855) & (0.913, 0.852) & (0.891, 0.847)
& (0.824, 0.904) & (0.917, 0.898) & (0.910, 0.943)\\
(1500,200) & (0.842, 0.865) & (0.922, 0.896) & (0.901, 0.835)
& (0.899, 0.845) & (0.880, 0.934) & (0.943, 0.913)\\
\midrule
(2000,50) & (0.929, 0.849) & (0.913, 0.886) & (0.899, 0.841)
& (0.897, 0.896) & (0.924, 0.934) & (0.897, 0.943)\\
(2000,100) & (0.903, 0.835) & (0.886, 0.882) & (0.879, 0.865)
& (0.926, 0.893) & (0.918, 0.932) & (0.925, 0.918)\\
(2000,150) & (0.913, 0.853) & (0.861, 0.890) & (0.916, 0.855)
& (0.863, 0.937) & (0.912, 0.938) & (0.938, 0.905)\\
(2000,200) & (0.859, 0.881) & (0.907, 0.906) & (0.923, 0.831)
& (0.913, 0.915) & (0.950, 0.929) & (0.899, 0.954)\\
\midrule
\midrule
\multirow{2}{*}{(Size, Overlap)} &\multicolumn{3}{c|}{MITRE ATT\&CK ICS} & \multicolumn{3}{c}{Threat Report} \\
& \texttt{Character} & \texttt{RecurChar} & \texttt{TokenText} & \texttt{Character} & \texttt{RecurChar} & \texttt{TokenText} \\
\midrule
(500,50) & (0.923, 0.912) & (0.903, 0.922) & (0.901, 0.918)
& (0.842, 0.869) & (0.867, 0.856) & (0.925, 0.921)\\
(500,100) & (0.918, 0.915) & (0.893, 0.911) & (0.880, 0.925)
& (0.869, 0.890) & (0.915, 0.853) & (0.936, 0.942)\\
(500,150) & (0.920, 0.916) & (0.952, 0.888) & (0.887, 0.932)
& (0.862, 0.869) & (0.930, 0.863) & (0.880, 0.930)\\
(500,200) & (0.918, 0.876) & (0.902, 0.910) & (0.929, 0.884)
& (0.858, 0.864) & (0.946, 0.860) & \textbf{(0.961, 0.948)} \\
\midrule
(1000,50) & (0.954, 0.905) & (0.927, 0.900) & (0.887, 0.929)
& (0.934, 0.915) & (0.910, 0.925) & (0.885, 0.858)\\
(1000,100) & (0.946, 0.935) & (0.932, 0.924) & (0.889, 0.929) & (0.937, 0.900) & (0.913, 0.921) & (0.847, 0.833)\\
(1000,150) & (0.962, 0.922) & (0.901, 0.927) & (0.870, 0.946) & (0.923, 0.875) & (0.920, 0.912) & (0.844, 0.932)\\
(1000,200) & \textbf{(0.942, 0.958)} & (0.923, 0.886) & (0.887, 0.929) & (0.894, 0.934) & (0.944, 0.842) & (0.782, 0.806)\\
\midrule
(1500,50) & (0.894, 0.898) & (0.905, 0.922) & (0.896, 0.925) & (0.965, 0.812) & (0.957, 0.866) & (0.885, 0.781)\\
(1500,100) & (0.917, 0.886) & (0.925, 0.923) & (0.924, 0.913) & (0.898, 0.888) & (0.915, 0.918) & (0.823, 0.907)\\
(1500,150) & (0.918, 0.893) & (0.923, 0.933) & (0.922, 0.898) & (0.910, 0.862) & (0.925, 0.883) & (0.833, 0.849)\\
(1500,200) & (0.913, 0.898) & (0.911, 0.928) & (0.884, 0.932) & (0.894, 0.934) & (0.944, 0.842) & (0.782, 0.806)\\
\midrule
(2000,50) & (0.912, 0.880) & (0.901, 0.901) & (0.911, 0.901) & (0.791, 0.916) & (0.872, 0.879) & (0.832, 0.867)\\
(2000,100) & (0.904, 0.886) & (0.892, 0.904) & (0.884, 0.935) & (0.871, 0.890) & (0.797, 0.884) & (0.913, 0.839)\\
(2000,150) & (0.906, 0.908) & (0.927, 0.899) & (0.931, 0.888) & (0.929, 0.935) & (0.872, 0.841) & (0.903, 0.891)\\
(2000,200) & (0.920, 0.891) & (0.909, 0.891) & (0.908, 0.904) & (0.925, 0.898) & (0.870, 0.901) & (0.891, 0.842)\\
\bottomrule
\bottomrule
\end{tabular}}
\end{table*}

To optimize the chunking strategy for documents' page\_content, we utilize the Ragas library~\cite{es2023ragas} in conjunction with all-MiniLM\footnote{\scriptsize \url{https://ollama.com/library/all-minilm}} (for embedding) and LLaMA3:8B (for evaluation) to search the most suitable chunking size, overlap, and splitting method for each dataset. We use content \textit{precision} and \textit{recall} as the key metrics: 
\begin{itemize}
    \item Precision measures whether all relevant items retrieved by the model are ranked higher than the irrelevant items;
    \item Recall measures how much of the relevant content is retrieved based on the annotated answers and the retrieved context.
\end{itemize}
Both precision and recall are evaluated within the range $[0,1]$, where a higher score indicates better performance.
As the datasets contain a huge number of samples, for practical efficiency, we select a subset of $1,000$ samples from each dataset except threat reports\footnote{\scriptsize We use all collected threat reports for generating testset and evaluation.}, generate a testset of $50$ items, and conduct evaluations based on the subset and testset.
This might not result in the optimal chunking size, overlap, and splitter method, but is enough to get a reasonable and useful chunking strategy for our practical applications.
As shown in Table~\ref{table:chunking}, we test the following commonly used configurations:
\begin{itemize}
    \item chunk sizes: $\{500, 1000, 1500, 2000\}$;
    \item overlaps: $\{50, 100, 150, 200\}$; 
    \item splitters: \texttt{Character}, \texttt{RecursiveCharacter}, and \texttt{TokenText}.
\end{itemize}
Our objective is to achieve high precision and recall simultaneously, ensuring that the system retrieves as many relevant documents as possible while minimizing irrelevant content. 
From the experimental results in Table~\ref{table:chunking}, it is easy to see that using the \texttt{RecursiveCharacter} splitter with a chunk size of $500$ and an overlap of $100$ is the most effective strategy for the VARIoT vulnerabilities, offering the best trade-off between precision and recall. 
Similarly, we can choose the suitable chunking strategies for VARIoT exploits, MITRE ATT\&CK ICS, and threat reports. The details are illustrated in Table~\ref{tab:chunkres}.

\begin{table}[]
    \centering
    \caption{Optimized chunking strategy for VARIoT vulnerabilities, exploits, ICS, and threat reports. \texttt{RecurChar} is short for \texttt{RecursiveCharacter}.}
    \label{tab:chunkres}
    \resizebox{0.45\textwidth}{!}{
    \begin{tabular}{c|ccc}
    \toprule \toprule
     {Dataset} & {Size} & {Overlap} & {Splitter Method} \\ \midrule
     VARIoT Vulns. & $500$ & $100$ & \texttt{RecurChar}\\
     VARIoT Exps. & $1000$ & $150$ & \texttt{TokenText} \\
     ICS & $1000$ & $200$ & \texttt{Character} \\
     Threat Reports & $500$ & $200$ & \texttt{TokenText} \\
     \bottomrule \bottomrule
    \end{tabular}}
\end{table}

\subsection{LLMs-based Evaluation of Outputs}\label{sec:exp-llmeval}

\begin{table*}
\centering
\caption{Comparison of \chatiot\ with LLM alone method (LLM-A). The experimental results are for moderate LLMs: LLaMA3:8B, LLaMA3.1:8B, and GPT-4o-mini. We use LLaMA3:70B as the evaluator for all experiments.}\label{tab:exp_llmsmall}
\begin{tabular}{c|c|cc|cc|cc}
\toprule \toprule
\multirow{2}{*}{Role} & \multirow{2}{*}{Metric} & \multicolumn{2}{c|}{LLaMA3:8B} &\multicolumn{2}{c|}{LLaMA3.1:8B} & \multicolumn{2}{c}{GPT-4o-mini} \\
              &                 & \chatiot & LLM-A & \chatiot & LLM-A & \chatiot & LLM-A \\
\midrule
\multirow{4}{*}{Consumer} & Reliability & 4.10 ($+$0.11) & 3.99 & 4.50 ($+$0.80) & 3.70 & 4.70 ($+$0.80) & 3.90 \\
                  & Relevance & 4.90 ($+$0.65) & 4.25 & 4.90 ($+$0.80) & 4.10 & 5.00 ($+$1.00) & 4.00 \\
                  & Technical & 4.50 ($+$0.47) & 4.03 & 4.40 ($+$0.80) & 3.60 & 4.45 ($+$0.55) & 3.90 \\
                  & Friendliness & 4.30 ($+$0.30) & 4.00 & 4.70 ($+$0.90) & 3.80 & 4.90 ($+$1.00) & 3.90 \\
\midrule
\multirow{4}{*}{Security Analyst} & Reliability & 4.30 ($+$0.26) & 4.04 & 4.75 ($+$0.67) & 4.08 & 4.85 ($+$1.19) & 3.66 \\
                  & Relevance & 4.63 ($+$0.33) & 4.30 & 4.95 ($+$0.87) & 4.08 & 4.91 ($+$1.21) & 3.70 \\
                  & Technical & 4.47 ($+$0.23) & 4.24 & 4.78 ($+$0.79) & 3.99 & 4.89 ($+$1.18) & 3.71 \\
                  & Friendliness & 4.04 ($+$0.01) & 4.03 & 4.02 ($+$0.87) & 3.15 & 4.92 ($+$1.43) & 3.49 \\
\midrule
\multirow{4}{*}{Technical Officer} & Reliability & 4.45 ($+$0.63) & 3.82 & 4.45 ($+$0.37) & 4.08 & 4.80 ($+$0.76) & 4.04 \\
                   & Relevance & 4.78 ($+$0.73) & 4.05 & 4.65 ($+$0.47) & 4.18 & 4.90 ($+$0.80) & 4.10 \\
                   & Technical & 4.59 ($+$0.83) & 3.76 & 4.48 ($+$0.29) & 4.19 & 4.83 ($+$0.68) & 4.15 \\
                   & Friendliness & 3.92 ($+$0.44) & 3.48 & 4.22 ($+$0.27) & 3.95 & 4.67 ($+$0.62) & 4.05 \\
\midrule
\multirow{4}{*}{Developer} & Reliability & 4.43 ($+$0.63) & 3.80 & 4.40 ($+$0.10) & 4.30 & 4.80 ($+$0.79) & 4.01 \\ 
& Relevance & 4.66 ($+$0.83) & 3.83 & 4.60 ($+$0.30) & 4.30 & 4.86 ($+$0.84) & 4.02 \\
& Technical & 4.67 ($+$0.62) & 4.05 & 4.50 ($+$0.10) & 4.40 & 4.94 ($+$0.92) & 4.02 \\
& Friendliness & 4.01 ($+$0.09) & 3.92 & 3.90 ($-$0.50) & 4.40 & 4.18 ($+$0.48) & 3.70 \\
\midrule
\multirow{4}{*}{Trainer} & Reliability & 4.10 ($-$0.14) & 4.24 & 4.30 ($-$0.19) & 4.49 & 4.55 ($+$0.29) & 4.26 \\
& Relevance & 4.56 ($+$0.16) & 4.40 & 4.40 ($-$0.11) & 4.51 & 4.64 ($+$0.24) & 4.40 \\
& Technical & 4.11 ($+$0.07) & 4.04 & 4.42 ($-$0.07) & 4.49 & 4.58 ($+$0.36) & 4.22 \\
 & Friendliness & 4.02 ($-$0.03) & 4.05 & 4.08 ($-$0.48) & 4.56 & 4.33 ($+$0.11) & 4.22 \\
\bottomrule \bottomrule
\end{tabular}
\end{table*}

As there is no public Question-Answer dataset about IoT security and threat intelligence, we synthesize 50 common IoT security-related questions (10 questions for each kind of user).
To evaluate our improvements, we compare \chatiot's outputs with the answers generated by underlying LLM alone (denoted as LLM-A), which is not equipped with our introduced IoT data sources. 
And we let another LLM be the evaluator and measure the quality of outputs by four metrics: \textit{Reliability}, \textit{Relevance}, \textit{Technical}, and \textit{Friendliness} as follows:
\begin{itemize}
    \item Reliability: the trustworthiness and reliability of each answer, ensuring it is plausible and aligns with known IoT best practices and standards.
    \item Relevance: Assess how well the answer addresses the specific question and meets the user’s needs, considering their role and context in the IoT ecosystem.
    \item Technical: Judge the appropriateness and precision of technical language, including IoT research, standards, protocols, and relevant technical aspects. Ensure that the answer demonstrates a solid understanding of IoT technologies.
    \item Friendliness: Determine how easy the answer is to comprehend, focusing on clarity for the user’s role, and how well the answer provides actionable steps or solutions tailored to the user’s IoT security needs. 
\end{itemize}
All scores are in $[0,5]$, where $5$ is the highest. 
For each question, the answers generated by \chatiot\ and corresponding LLM-A are inputted into the evaluator simultaneously. 
This approach ensures that both answers for each question are evaluated within the same internal state of the evaluator. By doing so, we aim to reduce the impact of LLM randomness as much as possible and enable a fair comparison between \chatiot\ and LLM-A.
The prompt for evaluation is shown in Figure~\ref{fig:llmevalprompt}.

\begin{figure}
    \centering    
{\footnotesize
\begin{tikzpicture}
% Draw rounded rectangle with shadow
\node[rectangle, rounded corners, draw=black, fill=black!3!white, text width=0.43\textwidth, inner sep=12pt, align=left] (box) {
    \textbf{\texttt{Task}}: 
    You are an expert IoT security assistant. Your task is to evaluate the answers to a question posed by the user of \{role\}; \\
    \vspace{5pt}
    \textbf{\texttt{Question}}: \{question\};\\
    \vspace{5pt}
    \textbf{\texttt{Answers}}:
    \begin{enumerate}
        \item \chatiot\_answer: \{\chatiot\_answer\};
        \item LLM-A\_answer: \{LLM-A\_answer\};
    \end{enumerate}
   \vspace{5pt}
    \textbf{\texttt{Instructions}}:
    \begin{itemize}
        \item \textbf{Criteria}: \{The descriptions about Reliability, Relevance, Technical, and User-friendliness.\}
        \item \textbf{Score}: \romannumeral1) Provide a score for each answer across the five metrics above. Scores should range from 0 to 5, with 5 being the highest and 0 being the lowest. 
        \romannumeral2) Scores should reflect how well each answer meets the criteria, particularly in alignment with the user role's background and needs.
        \item \textbf{Output Format}: 
        \romannumeral1) Present a table that includes the names of all answers and their scores for each metric. You can score differently for different metrics. 
        \romannumeral2) Give explanations for scores and output should be in Markdown format in English by default.
    \end{itemize}
};

\end{tikzpicture}
}
\caption{The prompt template for LLM-based evaluation of outputs.}
\label{fig:llmevalprompt}
\end{figure}

\begin{table*}
\centering
\caption{Comparison of \chatiot\ with LLM alone method (LLM-A). The experimental results are for the most advanced LLMs LLaMA3.1:70B and GPT-4o. We use LLaMA3:70B as the evaluator for all experiments.}\label{tab:exp_llm}
\begin{tabular}{c|c|cc|cc}
\toprule \toprule
\multirow{2}{*}{Role} & \multirow{2}{*}{Metric} & \multicolumn{2}{c|}{LLaMA3.1:70B} & \multicolumn{2}{c}{GPT-4o} \\
&  & \chatiot & LLM-A & \chatiot & LLM-A \\
\midrule
\multirow{4}{*}{Consumer} & Reliability & 4.40 ($+$0.70) & 3.70 & 4.55 ($+$0.55) & 4.00 \\
& Relevance & 4.90 ($+$0.90) & 4.00 & 4.85 ($+$0.75) & 4.10 \\
& Technical & 4.50 ($+$0.70) & 3.80 & 4.40 ($+$0.25) & 4.15 \\
& Friendliness & 4.70 ($+$1.00) & 3.70 & 4.80 ($+$0.75) & 4.05 \\
\midrule
\multirow{4}{*}{Security Analyst} & Reliability & 4.70 ($+$0.65) & 4.05 & 4.80 ($+$1.29) & 3.51 \\
& Relevance & 4.93 ($+$0.81) & 4.12 & 4.83 ($+$1.18) & 3.65 \\
& Technical & 4.82 ($+$0.84) & 3.98 & 4.83 ($+$1.18) & 3.65 \\
& Friendliness & 4.09 ($+$0.80) & 3.29 & 4.14 ($+$0.61) & 3.45 \\
\midrule
\multirow{4}{*}{Technical Officer} & Reliability & 4.45 ($+$0.37) & 4.08 & 4.85 ($+$0.61) & 4.24 \\
& Relevance & 4.71 ($+$0.51) & 4.20 & 4.88 ($+$0.63) & 4.25 \\
& Technical & 4.55 ($+$0.31) & 4.24 & 4.88 ($+$0.65) & 4.23 \\
& Friendliness & 4.13 ($+$0.27) & 3.86 & 4.49 ($+$0.61) & 3.88 \\
\midrule
\multirow{4}{*}{Developer} & Reliability & 4.35 ($-$0.14) & 4.49 & 4.75 ($+$0.93) & 3.82 \\
& Relevance & 4.54 ($+$0.02) & 4.52 & 4.81 ($+$0.97) & 3.84 \\
& Technical & 4.52 ($+$0.02) & 4.52 & 4.86 ($+$0.87) & 3.99 \\
& Friendliness & 3.96 ($-$0.24) & 4.30 & 4.11 ($+$0.63) & 3.48 \\
\midrule
\multirow{4}{*}{Trainer} & Reliability & 4.30 ($-$0.31) & 4.61 & 4.40 ($+$0.17) & 4.23 \\
& Relevance & 4.62 ($-$0.08) & 4.70 & 4.54 ($+$0.39) & 4.15 \\
& Technical & 4.31 ($-$0.23) & 4.54 & 4.53 ($+$0.21) & 4.32 \\
& Friendliness & 4.25 ($-$0.40) & 4.65 & 4.46 ($+$0.22) & 4.24 \\
\bottomrule \bottomrule
\end{tabular}
\end{table*}

We develop five versions of \chatiot\ using LLaMA3:8B, LLaMA3.1:8B, LLaMA3.1:70B, GPT-4o-mini, and GPT-4o, and employ LLaMA3:70B to evaluate them.
Table~\ref{tab:exp_llmsmall} shows the experimental results for moderate LLMs LLaMA3:8B, LLaMA3.1:8B, and GPT-4o-mini; and Table~\ref{tab:exp_llm} present the results for more advanced LLMs LLaMA3.1:70B and GPT-4o.
We also compute our improved scores over LLM-A in the tables.
From these results, several key observations can be made: 
\begin{itemize}
\item \chatiot\ significantly enhances the moderate LLM's performance in the IoT security domain. As shown in Table~\ref{tab:exp_llmsmall}, \chatiot\ achieves higher scores across most metrics for use cases \textit{Consumer}, \textit{Security Analyst}, \textit{Technical Officer}, and \textit{Developer}.
This is expected, as \chatiot\ integrates domain-specific IoT security knowledge, \eg, vulnerabilities and exploits, and tailors responses to be more user-friendly and relevant.
As illustrated in Table~\ref{tab:exp_llm}, \chatiot\ can also improve the advanced LLMs' capabilities in IoT problems.

\item However, \chatiot\ does not always outperform the baseline LLMs.
Taking the use case \textit{Trainer}, when using LLaMA series models, \chatiot\ even performs slightly worse; when using GPT-4o-mini and GPT-4o, the improvements achieved by \chatiot\ are much less than that for the other use cases.
This is likely due to the external data introduced in \chatiot\ focusing mainly on vulnerabilities, exploits, and TTPs, while lacking sufficient information on course training materials.
As a result, \chatiot\ excels at producing technical and security-centric content but may overlook broader aspects like training programs.
\end{itemize}

The above analysis also highlights the importance of incorporating external knowledge to bolster LLMs in specialized domains.
Fortunately, additional information, such as training materials, can easily be integrated into \chatiot\ using our \datakit\ toolkit.

\subsection{Analysis of Human Evaluation}\label{sec:exp-human}

\begin{figure}
    \centering
    \includegraphics[width=\linewidth]{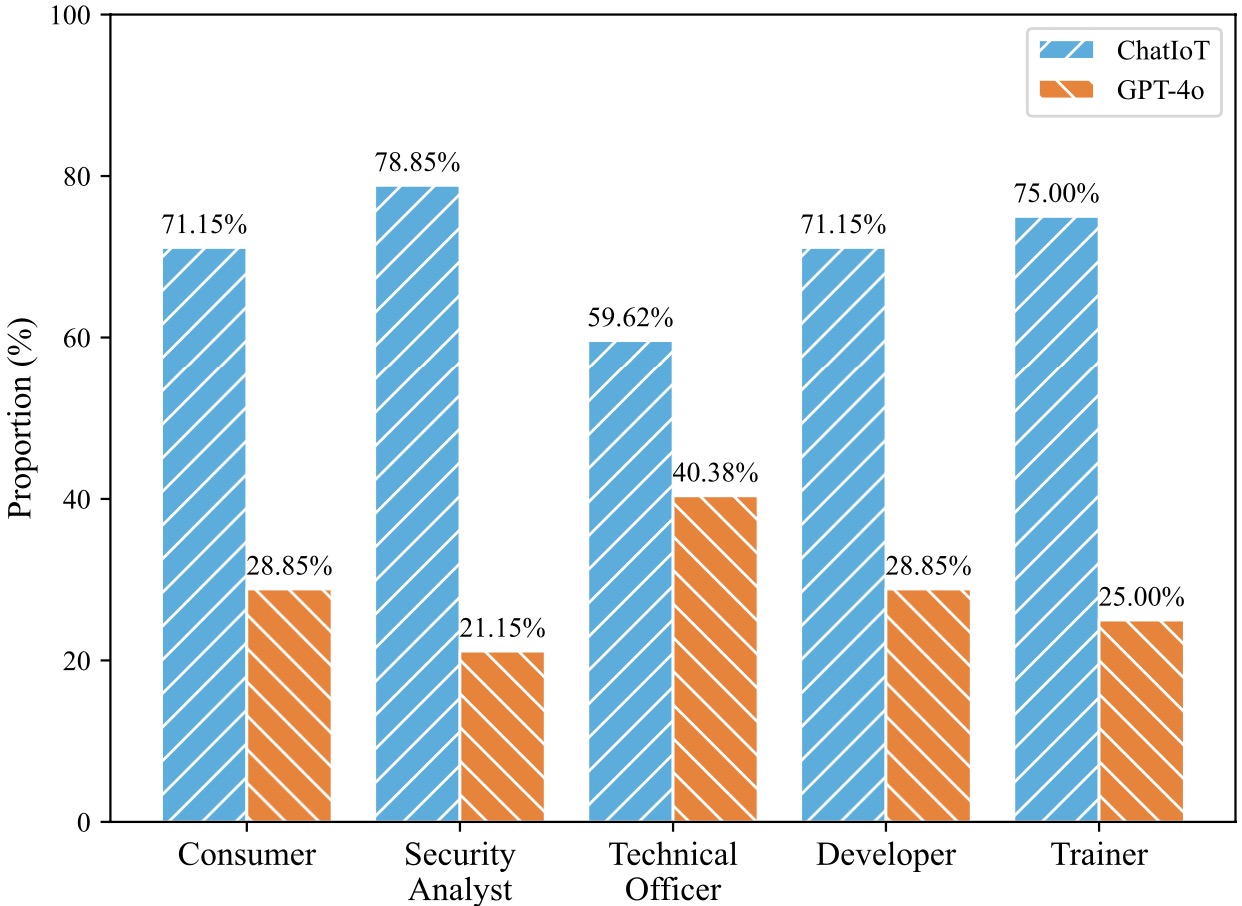}
    \caption{The human evaluation of \chatiot\ and LLM-A. The LLM of \chatiot\ and LLM-A method is GPT-4o in experiments.}
    \label{fig:human-exp}
\end{figure}

In Figure~\ref{fig:human-exp}, we present the results of human evaluations comparing \chatiot\ with GPT-4o, noting that \chatiot\ is built on top of GPT-4o in this experiment. We select four Q\&A pairs for each user type and display them as a survey. 
Human evaluators are asked to select the answer they find more suitable. 
To aid in their decisions, we provide them with evaluation metrics (as described in \S~\ref{tab:exp_llmsmall}), but do not require them to score each metric. 
This approach is intended to streamline the evaluation process, making it more akin to real-world scenarios where users prioritize ease of decision-making.

From the experimental results, it is clear that \chatiot\ consistently outperforms GPT-4o across all use cases. This aligns with expectations, as \chatiot\ integrates additional IoT-specific intelligence into the LLM. Notably, \chatiot\ demonstrates the greatest improvement for Security Analyst and the least for Technical Officer. The former result aligns with Table~\ref{tab:exp_llm}, where the comparison for Security Analyst shows the most significant difference.
Additionally, we give a use case comparison for \textit{Security Analyst} between GPT-4o and GPT-4o-based \chatiot. The formal analysis can be referred to Appendix~\ref{appendix:use-case-cmp}.
However, while Table~\ref{tab:exp_llm} suggests that the least improvement is for Trainer, the human evaluation indicates Technical Officer experiences the smallest gains. This discrepancy can be explained by: \romannumeral1) The improvements for Technical Officer, though better than those for Trainer, particularly in top metric scores, may not be as easily discernible to humans as other use cases, making it harder for them to identify notable differences; \romannumeral2) The Q\&A tasks for Technical Officer are generally more complex and technical than those for Trainer, making it easier for users to select a better answer in Trainer case even when improvements are less significant.

We acknowledge that our Q\&A survey might carry some biases since it is impossible to cover all use cases and questions/queries. However, we have made a strong effort to choose the most common and representative Q\&A pairs for this survey, which we believe adequately reflects the improvements achieved.
Nonetheless, \chatiot\ consistently provides better results over GPT-4o alone in the IoT security domain, demonstrating its enhanced capability to address IoT-specific challenges.

%Faithfulness: This measures the factual consistency of the generated answer against the given context. It is calculated from answer and retrieved context. The answer is scaled to (0,1) range. Higher the better. (only ChatIoT) 

%Context Relevance (ConRel) (only ChatIoT)

%answer\_relevancy: measures how relevant the generated response is to the given question (ChatIoT and baselines)

%llm\_grader: leverages a general-purpose zero-shot prompt to rate responses from an LLM to a given question on a scale from 1-10; (ChatIoT and baselines) 

%Models: llama3:8b

\section{Related Work}\label{sec:relatedwork}

Internet of Things (IoT) has seen rapid advancements in recent years, becoming an integral part of various domains, such as smart industries and homes, and serving as a key enabler in modern society.
However, despite its growth, IoT continues to face numerous security challenges, prompting significant research efforts aimed at improving IoT security.
With the rise of artificial intelligence (AI), machine learning (ML) and deep learning (DL)-based approaches have become increasingly popular in designing defense mechanisms for IoT devices, including malicious traffic classification~\cite{luo2022transformer,shafiq2020corrauc}, malware detection~\cite{vasan2020mthael,chaganti2022deep,aung2022atlas}, vulnerability discovery~\cite{neshenko2019demystifying}, and others~\cite{al2020survey,otoum2022dl,tambe2019detection}.

More recently, inspired by the success of large language models (LLMs), researchers have begun exploring the potential of LLMs to enhance IoT-related security tasks.
For instance, LLMs have been applied to existing IoT security challenges such as threat detection and fuzzing. Ferrag \etal~\cite{sokiotllm} introduced a BERT-based model, SecurityBERT, to achieve better cyber threat detection accuracy over traditional ML and DL-based methods. 
Similarly, Ma \etal~\cite{ma} and Wang \etal~\cite{llmiotfuz} proposed LLM-assisted fuzzing methods to uncover hidden bugs in IoT devices, enabling the detection of complex vulnerabilities that traditional techniques might miss.
Additionally, Yang \etal~\cite{yang2023iot} combined LLMs with static code analysis using prompt engineering to create a cost-effective solution for IoT vulnerability detection.
\cite{ji2024sevenllm} collected cybersecurity raw texts to train cybersecurity LLM to augment the analysis of cybersecurity events, and \cite{llmtikg} made use of a larger LLM to build knowledge graphs from public threat intelligence and use GPT to create datasets to fine-tune a smaller LLM to extract entities and TTPs from attack description.
Ferraris \etal~\cite{ferraris2024ici} proposed utilizing ChatGPT to enhance IoT trust semantics, aligning with W3C Web of Things (WoT) recommendations\footnote{\scriptsize \url{https://www.w3.org/WoT/}}.
This work extends the TrUStAPIS framework~\cite{ferraris2020trustapis}.

Beyond the above tasks, LLMs have been employed in other IoT challenges.
Meyuhas \etal~\cite{meyuhas2024iotlabel} used LLMs to address the problem of labeling previously unseen IoT devices.
\cite{llmiotcontrol,cui2024llmind} explored leveraging LLMs to control IoT devices and facilitate effective collaboration among them.
Mo \etal~\cite{mo2024iot} collected IoT sensor-natural language paired data and trained IoT-LM to interpret and interact with physical IoT sensors.
Xu \etal~\cite{xu2024penetrative} employed ChatGPT to interpret IoT sensor data and reason over tasks in the physical realm, introducing novel ways of integrating human knowledge into cyber-physical systems. 

Recently, Deldari \etal~\cite{deldari2024auditnet} proposed AuditNet, a conversational AI-based security assistant, which is most similar to \chatiot\ and also augmented by external knowledge.
However, AuditNet focused on standards, policies, and regulations of portable document format (PDF), and aimed to reduce the manual effort of security experts involved in compliance checks of IoT. 
On the other hand, we integrate IoT threat intelligence of various sources into \chatiot\ and can assist multiple kinds of users. Besides, we provide an end-to-end toolkit to process data in various formats, not limited to PDF. 

Together, these studies indicate that LLMs have great potential to improve the security of IoT systems in various domains, from vulnerability discovery to trustworthiness management. 
By integrating LLMs with IoT-specific threat intelligence, these models can be guided to meet the unique challenges posed by the IoT ecosystem.
Moreover, the continuous advancements in the LLM community, combined with increasingly accessible IoT datasets, are likely to further drive the adoption of LLMs in IoT-related research and practical applications.

\section{Conclusion}\label{sec:con}
In this work, we propose \chatiot, an LLM-based IoT security assistant, and conduct extensive evaluations on several common use cases. 
Concretely, we leverage the advanced language understanding and reasoning capabilities of LLM and IoT security and threat information to provide IoT security assistance, and develop an easy-to-use data processing toolkit. 
With our proposed LLM-based generation system and developed toolkit, \chatiot\ is easily scalable to integrate different kinds of IoT threat intelligence from multiple sources.

\smallskip
\noindent \textbf{Limitations \& Future Work.}
We acknowledge the limitations of our work and propose several potential directions for future improvements.
Firstly, although we have developed a data processing toolkit to handle various IoT security datasets, gathering these datasets from public sources on the Internet remains largely manual.
This process is highly dependent on human expertise, especially in identifying relevant data sources. 
To address this, a key future direction would be the design or integration of an automated library within \chatiot\ that could continuously update and ingest the latest or real-time IoT threat intelligence, significantly reducing the manual efforts involved.
Another promising direction involves re-training or fine-tuning an LLM specifically in the IoT security domain.
While our current implementation leverages a general-purpose LLM enhanced by external data sources, re-training or fine-tuning allows the LLM to more deeply understand the nuances and technical challenges unique to IoT security.
Such an approach could be integrated into \chatiot\ to provide even more reliable, technical, and actionable insights, pushing the boundaries of current IoT security solutions.
Together, these enhancements would make \chatiot\ not only more effective in processing and presenting IoT security and threat intelligence but also more capable of adapting to the fast-evolving landscape of cybersecurity threats.

\section*{Acknowledgements} This research is supported by the National Research Foundation, Singapore, under its National Satellite of Excellence Programme “Design Science and Technology for Secure Critical Infrastructure: Phase II” (Award No: NRF-NCR25-NSOE05-0001). Any opinions, findings and conclusions or recommendations expressed in this material are those of the author(s) and do not reflect the views of National Research Foundation, Singapore.

\appendix

\subsection{Background Specifications}\label{appendix:bk}
We present the background specifications for Security Analyst, Technical Officer, Developer, and Trainer in Figure~\ref{fig:bk-sec}.
It is also easy to add more types of users by clearly defining their background specifications.

\begin{figure*}
    \centering    
{\footnotesize
\begin{tikzpicture}
% Draw rounded rectangle with shadow
\node[rectangle, rounded corners, draw=black, fill=black!3!white, text width=0.95\textwidth, inner sep=12pt, align=left] (box) {

\textbf{Background of Security Analyst}
\vspace{5pt}\\
    \textbf{\texttt{"Knowledge"}}: "Security Analyst is an expert in identifying vulnerabilities, analyzing threats, and ensuring IoT devices are secure from cyber threats. Security Analyst possesses in-depth technical knowledge of security protocols, vulnerabilities, and exploits, and are proficient in interpreting complex security data."
    \vspace{5pt}\\
    \textbf{\texttt{"Goals"}}: "The primary aim is to conduct in-depth analyses of IoT security threats, vulnerabilities, and exploits, contributing to the development of secure IoT systems, and provide deep insights into potential attack vectors, technical analysis, and mitigation strategies."
    \vspace{5pt}\\
    \textbf{\texttt{"Requirements"}}: "Security Analyst requires detailed information about the vulnerabilities, exploits, and technical configurations of IoT devices."
\vspace{10pt}\\

\textbf{Background of Technical Officer}
\vspace{5pt}\\
\textbf{\texttt{"Knowledge"}}: "Technical Officer is familiar with security patch management, ensuring devices adhere to organizational security standards, and handling technical troubleshooting. "
\vspace{5pt}\\
\textbf{\texttt{"Goals"}}: "Technical Officer is responsible for overseeing the implementation and maintenance of secure IoT systems within an organization, applying security patches, enforcing security policies, and troubleshooting security issues."
\vspace{5pt}\\
\textbf{\texttt{"Requirements"}}: "Technical Officer's focus is on implementing security measures within the organization's infrastructure. You need practical steps to deploy security updates and verify compliance with security standards."
\vspace{10pt}\\

\textbf{Background of Developer}
\vspace{5pt}\\
\textbf{\texttt{"Knowledge"}}: "Developer works on the technical design and architecture of IoT devices, with a focus on incorporating security into product design. Developer has a deep understanding of device security, encryption protocols, and compliance with security regulations."
\vspace{5pt}\\
\textbf{\texttt{"Goals"}}: "Developer is responsible for ensuring IoT products meet industry security standards and are resilient against known threats, and designing and developing secure IoT devices."
\vspace{5pt}\\
\textbf{\texttt{"Requirements"}}: "Developer needs insights into current vulnerabilities, designs best practices, and how to avoid common security pitfalls in future product iterations."
\vspace{10pt}\\
\textbf{Background of Trainer}
\vspace{5pt}\\
\textbf{\texttt{"Knowledge"}}: "Trainer creates educational material or conducts training sessions to teach IoT security to a broader audience, including technical and non-technical participants. Trainer understands both technical and pedagogical aspects of IoT security and can explain complex concepts in a simplified manner."
\vspace{5pt}\\
\textbf{\texttt{"Goals"}}: "Trainer aims to guide others in the best practices for IoT security, helping to raise awareness and improve security practices across different user groups."
\vspace{5pt}\\
\textbf{\texttt{"Requirements"}}: "Trainer needs information that can be used in a training environment, with clear examples, case studies, and simplified explanations for different levels of learners."
};
\end{tikzpicture}
}
\caption{The background specifications for Security Analyst, Technical Officer, Developer, and Trainer, utilized to guide the \chatiot\ to generate answers.}
\label{fig:bk-sec}
\end{figure*}

\subsection{Metadata Information and Examples}\label{appendix:metadataother}

\begin{figure*}
    \centering
    \includegraphics[width=\linewidth]{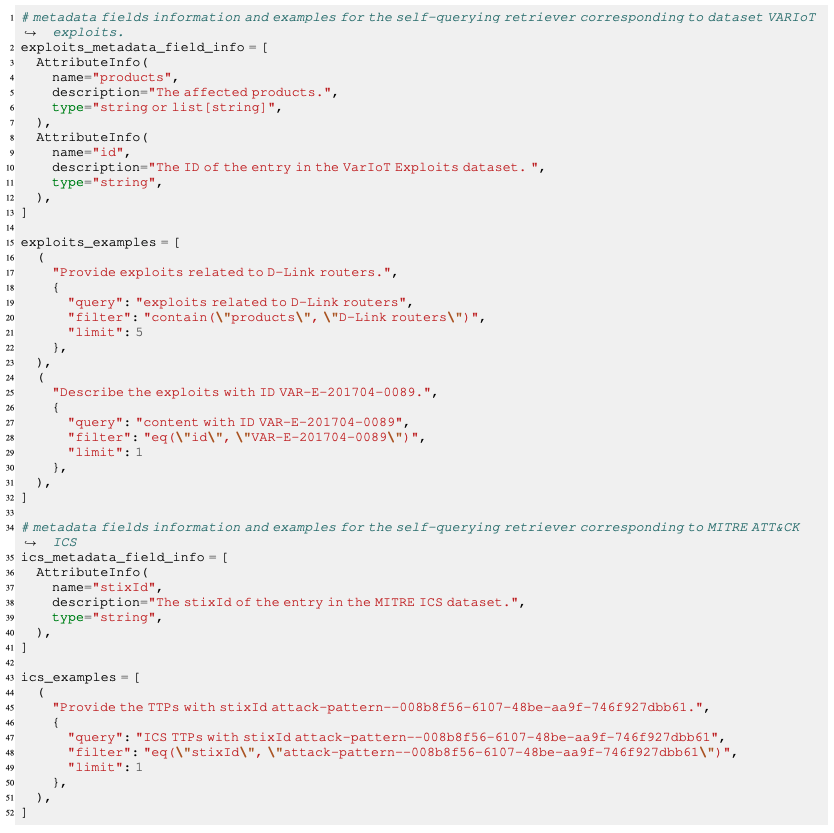}
    \caption{The metadata field information and examples for VARIoT exploits and MITRE ATT\&CK ICS.}
\label{fig:metaothers}
\end{figure*}

\begin{figure*}
    \centering
    \includegraphics[width=\linewidth]{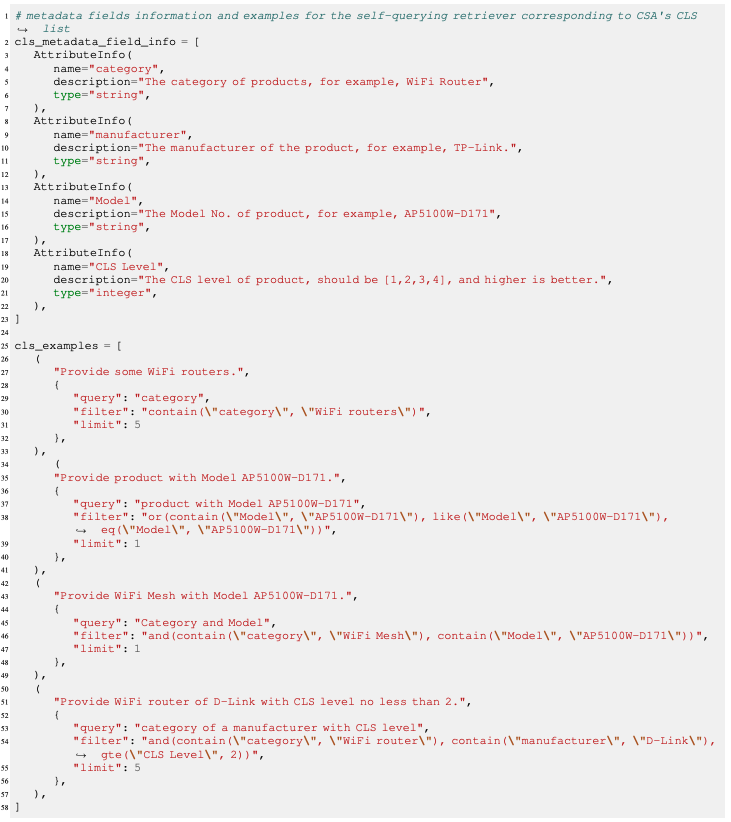}
    \caption{The metadata field information and examples for CLS.}
\label{fig:metaothers2}
\end{figure*}

We present the metadata information and examples for VATIoT exploits, MITRE ATT\&CK ICS, and CLS are shown in Figure~\ref{fig:metaothers} and~\ref{fig:metaothers2}.

\subsection{Use Case Study: Security Analyst}\label{appendix:use-case-cmp}

We compare the generated outputs of \chatiot\ to that of GPT-4o to show our benefits from external IoT security and threat information and knowledge.
In Fig.~\ref{fig:eva-cmp}, we present the outputs for the query \textit{“Investigate the vulnerabilities and exploits associated with TP-Link routers”} from the perspective of a \textit{Security Analyst}. The key advantages of \chatiot\ over GPT-4o are summarized as follows:

\begin{itemize}
    \item \textbf{Reliability}: GPT-4o provides general vulnerabilities and examples but lacks specific details, such as CVE references for many vulnerabilities. It misses certain critical vulnerabilities. On the other hand, \chatiot\ offers precise information about vulnerabilities, including affected models and specific risks. It references real-world CVEs, enhancing factual reliability.
    
    \item \textbf{Relevance}: GPT-4o covers a broad range of vulnerabilities but does not specifically target the needs of a security analyst, making it less relevant for professionals. 
    \chatiot's answer is tailored for a security analyst, focusing on vulnerabilities impacting network security. It provides detailed exploit analysis and mitigation strategies, making it highly relevant.

    \item \textbf{Technical}: GPT-4o is a general LLM, so it lacks comprehensive technical details about each vulnerability and does not explain them in depth or provide mitigation strategies.
    \chatiot\ offers detailed technical specifics for each vulnerability, including descriptions, risks, and mitigation strategies. The structured analysis of exploits is suitable for technical audiences.

    \item \textbf{Friendliness}: GPT-4o uses straightforward language, making it easy for a general audience to understand, but lacks engagement for detailed insights.
    Our approach maintains a professional tone while being informative. The structured layout enhances readability, making it user-friendly for professionals seeking specific information.
\end{itemize}

In a nutshell, \chatiot\ outperforms using only GPT-4o to processing IoT security-related queries/questions in reliability, relevance, and technical depth, making it more suitable for a security analyst's needs. While the GPT-4o's answer is friendly and accessible, it lacks the detailed information and focus required for professionals in IoT security.

\begin{figure*}
    \centering
\begin{tikzpicture}
[node distance=0.1cm, box/.style={rectangle, rounded corners, fill=blue!5, text width=\textwidth, minimum height=1cm, align=left}, title/.style={font=\large\bfseries}]

\node[box, fill=green!5] (GPT4o){
\tiny
\noindent TP-Link routers, like many other networking devices, have been subject to various vulnerabilities and exploits over the years. These vulnerabilities can range from simple misconfigurations to more complex software flaws that can be exploited by attackers. Here are some notable vulnerabilities and exploits associated with TP-Link routers:

\noindent \textbf{1. Default Credentials and Weak Passwords}

Many TP-Link routers come with default administrative credentials (e.g., admin/admin). If these credentials are not changed, attackers can easily gain access to the router's administrative interface. Weak passwords can also be brute-forced.

\noindent \textbf{2. Remote Code Execution (RCE)}

Several TP-Link router models have been found to have vulnerabilities that allow remote code execution. For example:
\begin{itemize}
    \item \textbf{CVE-2019-7406}: A vulnerability in the TP-Link WR940N allows remote attackers to execute arbitrary code via a crafted HTTP request.
    \item \textbf{CVE-2020-12109}: A buffer overflow vulnerability in the TP-Link Archer C5 allows remote attackers to execute arbitrary code.
\end{itemize}

\noindent \textbf{3. Cross-Site Scripting (XSS)}

Cross-Site Scripting vulnerabilities can allow attackers to inject malicious scripts into the web interface of the router. For example:
\begin{itemize}
    \item \textbf{CVE-2019-7405}: An XSS vulnerability in the TP-Link WR940N allows remote attackers to inject arbitrary web script or HTML.
\end{itemize}

\noindent \textbf{4. Cross-Site Request Forgery (CSRF)}

CSRF vulnerabilities can allow attackers to perform actions on behalf of the authenticated user without their consent. For example:
\begin{itemize}
    \item \textbf{CVE-2018-11714}: A CSRF vulnerability in the TP-Link Archer C1200 allows remote attackers to change the router's settings.
\end{itemize}

\noindent \textbf{5. Command Injection}

Command injection vulnerabilities allow attackers to execute arbitrary commands on the router's operating system. For example:
\begin{itemize}
    \item \textbf{CVE-2017-13772}: A command injection vulnerability in the TP-Link WR1043ND allows remote attackers to execute arbitrary commands via the ping feature.
\end{itemize}

\noindent \textbf{6. Information Disclosure}

Information disclosure vulnerabilities can expose sensitive information about the router or the network. For example:
\begin{itemize}
    \item \textbf{CVE-2019-7404}: An information disclosure vulnerability in the TP-Link WR940N allows remote attackers to obtain sensitive information via a crafted HTTP request.
\end{itemize}

\noindent \textbf{7. Firmware Vulnerabilities}

Outdated firmware can contain numerous vulnerabilities that have been patched in later versions. It's crucial to keep the router's firmware up to date to mitigate these risks.

\noindent \textbf{8. Misconfigurations}

Improper configurations, such as enabling remote management without proper security measures, can expose the router to external attacks.

\noindent \textbf{Mitigation Strategies}

To protect against these vulnerabilities, consider the following best practices:
\begin{itemize}
    \item \textbf{Change Default Credentials}: Always change the default username and password to something strong and unique.
    \item \textbf{Update Firmware}: Regularly check for and apply firmware updates from TP-Link.
    \item \textbf{Disable Unnecessary Services}: Turn off services like remote management if they are not needed.
    \item \textbf{Use Strong Encryption}: Ensure that Wi-Fi networks use strong encryption methods like WPA3.
    \item \textbf{Network Segmentation}: Segment your network to limit the impact of a compromised device.
    \item \textbf{Monitor Network Traffic}: Use network monitoring tools to detect unusual activity.
\end{itemize}

\noindent \textbf{Conclusion}

TP-Link routers, like any other network devices, can be vulnerable to various types of attacks if not properly secured. Regular updates, strong passwords, and proper configuration can significantly reduce the risk of exploitation. Always stay informed about the latest vulnerabilities and apply necessary patches and mitigations promptly.
};
\node[box, below=of GPT4o] (ChatIoT-gpt4o) {
\tiny
\begin{center}
\textbf{Technical Analysis of Vulnerabilities and Exploits in TP-Link Routers}
\end{center}

\noindent \textbf{Overview}

TP-Link routers have been identified with multiple security vulnerabilities that pose significant risks to users. This analysis focuses on the vulnerabilities, associated risks, and actionable security measures to mitigate these threats.

\noindent \textbf{Identified Vulnerabilities}

\begin{enumerate}
    \item \textbf{Denial of Service (DoS) Vulnerability}
    \begin{itemize}
        \item \textbf{Affected Models}: TP-Link Archer C2, Archer C20i
        \item \textbf{Description}: This vulnerability allows attackers to disrupt the normal functioning of the router, rendering it unavailable to legitimate users.
        \item \textbf{Risk}: High. A successful DoS attack can lead to network downtime, affecting all connected devices.
        \item \textbf{Mitigation}: Implement rate limiting and anomaly detection to identify and block malicious traffic. Regularly update firmware to patch known vulnerabilities.
    \end{itemize}

    \item \textbf{Security Bypass Vulnerability}
    \begin{itemize}
        \item \textbf{Affected Models}: TP-Link Archer C2, Archer C20i
        \item \textbf{Description}: This vulnerability allows attackers to bypass security mechanisms, gaining unauthorized access to the router.
        \item \textbf{Risk}: High. Unauthorized access can lead to data breaches and further exploitation.
        \item \textbf{Mitigation}: Enable strong authentication mechanisms, such as multi-factor authentication (MFA). Regularly review and update access control lists (ACLs).
    \end{itemize}

    \item \textbf{Cross-Site Request Forgery (CSRF)}
    \begin{itemize}
        \item \textbf{Affected Models}: TP-Link TL-WR1043N
        \item \textbf{Description}: This vulnerability allows attackers to perform unauthorized actions on behalf of authenticated users by tricking them into clicking malicious links.
        \item \textbf{Risk}: Medium. CSRF can lead to unauthorized configuration changes, such as altering admin passwords.
        \item \textbf{Mitigation}: Implement anti-CSRF tokens in web forms. Educate users about the risks of clicking on unknown links.
    \end{itemize}

    \item \textbf{Path Traversal Vulnerability}
    \begin{itemize}
        \item \textbf{Affected Models}: TP-Link TL-R600VPN
        \item \textbf{Description}: This vulnerability allows attackers to access sensitive system files by manipulating URL paths.
        \item \textbf{Risk}: High. Unauthorized access to system files can lead to information disclosure and further exploitation.
        \item \textbf{Mitigation}: Validate and sanitize user inputs to prevent path traversal attacks. Regularly update firmware to patch known vulnerabilities.
    \end{itemize}

    \item \textbf{Command Injection Vulnerability}
    \begin{itemize}
        \item \textbf{Affected Models}: TP-Link Archer C2, Archer C20i
        \item \textbf{Description}: This vulnerability allows attackers to execute arbitrary commands on the router.
        \item \textbf{Risk}: Critical. Command injection can lead to complete system compromise.
        \item \textbf{Mitigation}: Implement input validation and sanitization. Use security mechanisms like Web Application Firewalls (WAF) to detect and block malicious inputs.
    \end{itemize}
\end{enumerate}

\noindent \textbf{Exploits Analysis}

\begin{enumerate}
    \item \textbf{Proof-of-Concept (PoC) Exploits}
    \begin{itemize}
        \item \textbf{Description}: Researchers have developed PoCs to demonstrate the feasibility of exploiting these vulnerabilities.
        \item \textbf{Risk}: High. PoCs can be used by malicious actors to develop real-world attacks.
        \item \textbf{Mitigation}: Monitor security advisories and apply patches promptly. Conduct regular security assessments to identify and mitigate potential vulnerabilities.
    \end{itemize}

    \item \textbf{Unauthenticated Reflected XSS}
    \begin{itemize}
        \item \textbf{Affected Models}: TP-Link TD-W8951ND
        \item \textbf{Description}: This vulnerability allows attackers to inject malicious scripts into the browser of an unauthenticated user.
        \item \textbf{Risk}: Medium. XSS can lead to session hijacking and unauthorized actions.
        \item \textbf{Mitigation}: Implement Content Security Policy (CSP) and sanitize user inputs. Regularly update firmware to patch known vulnerabilities.
    \end{itemize}

    \item \textbf{Authenticated Reflected XSS}
    \begin{itemize}
        \item \textbf{Affected Models}: TP-Link TD-W8951ND
        \item \textbf{Description}: This vulnerability allows attackers to inject malicious scripts into the browser of an authenticated user.
        \item \textbf{Risk}: Medium. XSS can lead to session hijacking and unauthorized actions.
        \item \textbf{Mitigation}: Implement CSP and sanitize user inputs. Educate users about the risks of clicking on unknown links.
    \end{itemize}

    \item \textbf{Reset Admin Password CSRF}
    \begin{itemize}
        \item \textbf{Affected Models}: TP-Link TD-W8951ND
        \item \textbf{Description}: This vulnerability allows attackers to reset the admin password by tricking the administrator into clicking a malicious link.
        \item \textbf{Risk}: High. Unauthorized access to the admin account can lead to complete system compromise.
        \item \textbf{Mitigation}: Implement anti-CSRF tokens and educate users about the risks of clicking on unknown links.
    \end{itemize}
\end{enumerate}

\noindent \textbf{Actionable Security Measures}

\begin{enumerate}
    \item \textbf{Firmware Updates}
Regularly update router firmware to patch known vulnerabilities.
Enable automatic updates if available.

    \item \textbf{Strong Authentication}
Use strong, unique passwords for admin accounts.
Enable multi-factor authentication (MFA) where possible.

    \item \textbf{Network Segmentation}
Segment the network to limit the impact of a compromised device.
Use VLANs to isolate sensitive devices.

    \item \textbf{Monitoring and Logging}
Implement network monitoring to detect and respond to suspicious activities.
Enable logging to maintain an audit trail of administrative actions.

    \item \textbf{User Education}
Educate users about the risks of phishing and social engineering attacks.
Encourage users to report suspicious activities.
\end{enumerate}

\noindent \textbf{Disclaimer}

The recommendations provided in this report are based on the information available at the time of writing. We are not legally responsible for any actions taken based on these recommendations. Always consult with a professional security expert before implementing any security measures.

};

\end{tikzpicture}

    \caption{Generated answers of GPT-4o (above) and GPT-4o-based \chatiot\ (below) for the inputs (User Role=Security Analyst, Query="Investigate the vulnerabilities and exploits associated with the TP-Link Routers").}
    \label{fig:eva-cmp}
\end{figure*}

\end{document}